\documentclass[apj]{emulateapj}
\usepackage{amsmath}
\usepackage{graphicx}
\usepackage{booktabs,threeparttable}
\usepackage{hyperref}
\usepackage{multirow}
\usepackage{ulem}
\shorttitle{Change of Field$-$Gas Alignment at Gravity-Driven Alfv\'enic Transition}
\shortauthors{Chen,\ King, \& Li}
\begin{document}
\title{Change of Magnetic Field$-$Gas Alignment at Gravity-Driven Alfv\'enic Transition \\ in Molecular Clouds: Implications for Dust Polarization Observations }
\author{Che-Yu Chen, Patrick K.~King, and Zhi-Yun Li}
\affil{Department of Astronomy, University of Virginia, Charlottesville, VA 22901}

\begin{abstract}

Diffuse striations in molecular clouds are preferentially aligned with local magnetic fields whereas dense filaments tend to be perpendicular to them. When and why this transition occurs remain uncertain. 
To explore the physics behind this transition, we compute the histogram of relative orientation (HRO) between the density gradient and the magnetic field in 3D MHD simulations of prestellar core formation in shock-compressed regions within GMCs.
We find that, in the magnetically-dominated (sub-Alfv\'enic) post-shock region, the gas structure is preferentially aligned with the local magnetic field. For overdense sub-regions with super-Alfv\'enic gas, their elongation becomes preferentially perpendicular to the local magnetic field instead. 
The transition occurs when self-gravitating gas gains enough kinetic energy from the gravitational acceleration to overcome the magnetic support against the cross-field contraction, which results in a power-law increase of the field strength with density.
Similar results can be drawn from HROs in projected 2D maps with integrated column densities and synthetic polarized dust emission.
We quantitatively analyze our simulated polarization properties, and interpret the reduced polarization fraction at high column densities as the result of increased distortion of magnetic field directions in trans- or super-Alfv\'enic gas. 
Furthermore, we introduce measures of the inclination and tangledness of the magnetic field along the line of sight as the controlling factors of the polarization fraction.
Observations of the polarization fraction and angle dispersion can therefore be utilized in studying local magnetic field morphology in star-forming regions.

\end{abstract}
\keywords{ISM: clouds -- ISM: magnetic fields -- magnetohydrodynamics (MHD) -- polarization -- stars: formation -- turbulence}

\section{Introduction}
\label{sec:intro}

Magnetic fields are widely believed to affect star forming processes at all physical scales and throughout different evolutionary stages in giant molecular clouds (GMCs), in combination with turbulence and gas self-gravity \citep{1987ARA&A..25...23S,2007ARA&A..45..565M}. The cloud-scale magnetic field can limit compression in turbulence-generated interstellar shocks that create dense clumps and filaments \citep{1956MNRAS.116..503M,1992phas.book.....S}, and sufficiently strong magnetic field may prohibit the formation or collapse of gravitationally bound cores \citep{1966MNRAS.132..359S,1976ApJ...210..326M}. On the other hand, the local magnetic field within collapsing cores can help remove angular momentum during the disk formation process \citep{1979MNRAS.187..311G,1985prpl.conf..320M,1991ApJ...373..169M,2003ApJ...599..363A,2014PPVI...Li}. The role played by magnetic fields during star formation within GMCs is therefore a complicated but important topic.

Observationally, the magnetic field strength along the line of sight can be directly measured via Zeeman splitting of molecular lines, but it is hard to detect \citep{1989ApJ...338L..61G,1993prpl.conf..279H,1999ApJ...520..706C,1999ApJ...514L.121C,2001ApJ...554..916B,2008A&A...487..247F}. 
In dense star-forming clouds, the polarized thermal emission from dust grains at infrared and sub-millimeter wavelengths has been shown to be a reliable technique of studying the projected magnetic field direction on the plane of sky \citep[e.g.][]{1984ApJ...284L..51H,1988QJRAS..29..327H,1997ApJ...487..320N,2000ApJS..128..335D,2000PASP..112.1215H,2001ApJ...562..400M}. Though the physical mechanism causing the alignment of grains with respect to the magnetic field is still an open question \citep[see e.g.][]{1996ApJ...470..551D,1997ApJ...480..633D,2000ASPC..215...69L,2003ApJ...589..289W,2005ApJ...631..361C,2007MNRAS.378..910L,2009ApJ...697.1316H}, it is generally recognized that elongated grains are oriented with their long axes perpendicular to the magnetic field lines, and thus the dust emission is linearly polarized perpendicular to the magnetic field (\citealt{1951ApJ...114..206D}; also see recent review of \citealt{2007JQSRT.106..225L}). The degree of polarization observed in infrared/sub-millimeter is typically $\sim 10~\%$ in interstellar clouds \citep[e.g.][]{2004A&A...424..571B,2015A&A...576A.105P,2016ApJ...824..134F} and $\sim 1~\%$ in dense cores \citep[e.g.][]{2000ApJ...537L.135W,2009ApJS..182..143M,2009ApJ...700..251T,2014ApJS..213...13H}.

Theoretically, magnetohydrodynamical (MHD) simulations offer an approach to calibrate the interpretation of polarimetric patterns. 
In infrared and sub-millimeter wavelengths, simulated polarized emission can be directly calculated by integrating the radiative transfer equations for the Stokes parameters along the line of sight, with assumptions of grain properties and their alignment efficiencies \citep{1985ApJ...290..211L,1990ApJ...362..120W,2000ApJ...544..830F,2003LNP...614..252O}.
Synthetic polarimetric maps can be produced from the three-dimensional (3D) gas density and magnetic field information known at every voxel. Most theoretical studies have focused on inferring the magnetic field strength from the distribution of polarization angles \citep{2001ApJ...561..800H,2001ApJ...546..980O}, while others have concentrated on providing insight into the polarization$-$intensity correlation (\citealt{2001ApJ...559.1005P,2008ApJ...679..537F}, also see Section~\ref{sec:pN}).

In numerical studies of star formation, overdense fragments within GMCs are often formed due to in-cloud shocks and colliding gas flows from cloud-scale turbulence \citep{1985prpl.conf..201S,2007prpl.conf...63B}. The density structure and gravitational instability induced by supersonic turbulence under the magnetic effects inside GMCs are therefore crucial topics among many theoretical studies \citep[e.g.][]{2001ApJ...546..980O,2004ApJ...605..800L,2008ApJ...687..354N,2011ApJ...728..123K}.
It is noteworthy that, though turbulence can create a combination of shearing, diverging, and converging effects at all physical scales, it is those regions with large-scale convergent flows that will compress gas significantly and strongly alter the gravitational stability in the cloud \citep{2004RvMP...76..125M}.
Recent work conducted by \citet[][hereafter CO14 and CO15]{2014ApJ...785...69C,2015ApJ...810..126C} therefore adopted an idealized model, considering a local region inside a GMC where two large-scale supersonic flows collide. They found that, in the dense layer compressed by these oblique MHD shocks, the amplified magnetic field is relatively well-ordered and strong enough for gas to preferably flow along the field lines. It is not until gravitationally bound prestellar cores form that the magnetic field starts to be bent by gas self-gravity (see Figure~13 of \hyperlink{CO14}{CO14}). This anisotropic condensation in locally flat region has been validated by observations \citep{2015AAS...22525602M}, and can help explain the general feature of dense filaments perpendicular to the magnetic field in their immediate vicinities \citep[see review in][]{2014prpl.conf...27A}.

With advanced technologies and instruments, polarization observations have become a powerful tool to  evaluate theoretical models of the magnetic effects during cloud evolution and core formation (e.g.~\citealt{2013ApJ...770..151C,2013ApJ...768..159H,2014ApJ...792..116Z,2015ApJ...814L..28C,2016ApJ...819..159C,2016ApJ...821...41L,2016A&A...586A.136P}, also see \citealt{2014prpl.conf..101L} for a recent review).
To characterize gas structure with respect to the magnetic field morphology, \cite{2013ApJ...774..128S} considered the histogram of relative orientation (HRO) between the density gradient and the magnetic field in 3D MHD simulations and the column density gradient and the polarization angle in two-dimensional (2D) maps that are integrated along the line of sight. Their results suggest that the shape of HROs is determined by the gas density, and the initial magnetization level of the cloud is imprinted on the observable polarization morphology.
This statistical tool provides a new way to study the role played by magnetic field in star-forming clouds, and has been applied to polarization observations of several regions \citep[e.g.][]{2016A&A...586A.138P}.

Here, motivated by \cite{2013ApJ...774..128S}, we analyze the prestellar core-forming simulations from \hyperlink{CO15}{CO15} by computing the HRO, and investigate the physics controlling HRO shape. We found that the angle between the density gradient and the magnetic field is highly correlated with the significance of self-gravity in dense structures. Specifically, the HRO changes shape when gravity is strong enough to accelerate the gas to a super-Alfv\'enic speed. In addition, there are similar trends in the shape of HROs determined from the 2D projected polarization maps, suggesting that polarization observations can be a tool to directly constrain the magnetic field morphology and strength.

The outline of this paper is as follows. We describe in Section~\ref{sec:methods} our computational and mathematical methods, briefly reviewing the numerical simulations adopted in this work, the definition of the HRO, and the derivation of polarization fraction. HROs in 3D space are analyzed in Section~\ref{sec:BvsD}, including the choices of the segmentation (Section~\ref{sec:ncut}) and the significance of self-gravity relative to the magnetic support implied by the shape of HRO (Section~\ref{sec:gravity}). We discuss the field-density relationship in Section~\ref{sec:bn}. In Section~\ref{sec:pvsN}, we extend our analysis to HROs in projected 2D map, where we argue that the threshold value for a power-law polarization$-$column density relation is the key for different HRO shapes.
We further provide a quantitative way to estimate the polarization fraction from the magnetic field morphology (Section~\ref{sec:pfactors}), explaining how it changes during cloud evolution.
Section~\ref{sec:sum} summarizes our conclusions.

\section{Methods}
\label{sec:methods}

\subsection{Numerical Simulations}
The simulations we adopted are described in \hyperlink{CO14}{CO14} and \hyperlink{CO15}{CO15} and briefly summarized here. These fully 3D MHD simulations, conducted using {\it Athena} \citep{2008ApJS..178..137S}, are $1$~pc-wide boxes, where plane-parallel converging flows collide head-to-head, creating a locally flat, dense layer in the post-shock region. 
This setup is similar to the situation when large-scale ($\sim 10-10^2$~pc) turbulent flows in the molecular cloud locally collide, and by adopting a box size much smaller than the driven scale of the turbulent flow, the convergent flow can be considered as plane-parallel.
Seeded by a randomly perturbed velocity field with power spectrum ${v_k}^2 \propto k^{-4}$, filamentary structures and prestellar cores grow within the shocked layer. 
By conducting the simulation multiple times with different realization of perturbation field (i.e. sampling different $1\times 1\times 1$-pc regions within the same cloud), one can collect enough information for statistical analysis.
The cloud magnetic field in the pre-shock region is initially set to be oblique with respect to the shock. In the post-shock region, the field lines become roughly parallel to the dense layer due to shock compression (see Figure~3-5 in \hyperlink{CO14}{CO14}). 
All models are isothermal at $T = 10$~K ($c_s=0.2$~km/s) with an initial gas density $n_0=10^3$~cm$^{-3}$.

We chose the ideal-MHD models with different inflow Mach numbers (${\cal M} = 5$, $10$, and $20$) and fixed cloud magnetic field $B_0 = 10~\mu$G at $20^\circ$ to the inflow, as surveyed in \hyperlink{CO15}{CO15}.\footnote{Models in \hyperlink{CO14}{CO14} include the additional physics of ambipolar diffusion, but their resolution is not as high as the ideal MHD simulations of \hyperlink{CO15}{CO15}.} This parameter set covers different turbulent levels in the cloud and includes a range of number densities, magnetic field strengths, and plasma $\beta$ values in the post-shock region (Table~\ref{sumtable}; also see Table~1 in \hyperlink{CO15}{CO15} for an extended summary of model parameters and physical properties of the post-shock regions). 
Note that, though \hyperlink{CO15}{CO15} included 6 different realizations for each model to get better statistics on core properties, we only consider one realization for each model here since there are enough number of cells in one run to carry the HRO and polarization analysis.
To include regions at different evolutionary stages of the cloud from quiescent to core-forming condition, we consider the snapshot when the most evolved core starts to collapse in each simulation, $t_\mathrm{collapse} = t (n_\mathrm{max} \geq 10^7~\mathrm{cm}^{-3})$, as defined in \hyperlink{CO14}{CO14} and \hyperlink{CO15}{CO15}.

\begin{table*}[t]
\begin{center}
  \begin{threeparttable}
\caption{Summary of numerical models and best-fit results}
 \label{sumtable}
 \begin{tabular}{l|ccc|ccc|cccc}
\tableline\tableline
Model & $\beta_0$ & ${\cal M}_0$ & ${\cal M}_\mathrm{A,0}$ & $\beta_\mathrm{ps}$\tablenotemark{a} & ${\cal M}_\mathrm{rms,ps}$\tablenotemark{a} & ${\cal M}_\mathrm{A,ps}$\tablenotemark{a} & $\log_{10} n_\mathrm{tr}$ & $\kappa$\tablenotemark{b} in $B\propto n^\kappa$ & $\log_{10} N_\mathrm{tr}$ & $\alpha$\tablenotemark{b} in $p \propto N^{-\alpha}$ \\
\hline
M5 & 0.39 & 5 & 2.2 & 0.25 & 2.2 & 0.78 & 5.0 & $0.61\pm 0.03$ &  22.2 & $0.37\pm 0.09$ \\
M10 & 0.39 & 10 & 4.4 & 0.16 & 2.9 & 0.81 & 5.3 & $0.59\pm 0.03$ &  22.3 & $0.03\pm 0.01$\\
M20 & 0.39 & 20 & 8.8 & 0.09 & 4.0 & 0.84 & 5.5 & $0.28\pm 0.02$ &  22.6 & $0.00\pm 0.03$\\
\tableline\tableline
\end{tabular}
    \begin{tablenotes}
    \item $^\mathrm{a}$ The plasma $\beta$, gas Mach number ${\cal M}_\mathrm{rms}$, and Alfv{\'e}n Mach number ${\cal M}_\mathrm{A}$ (see Equation~(\ref{MAdef})) of the post-shock layer are measured at $t=0.2$~Myr to exclude gravitational effects from overdense regions. See \hyperlink{CO14}{CO14} and \hyperlink{CO15}{CO15} for justification.
    \item $^\mathrm{b}$ With $95\%$ confidence bounds.
    \end{tablenotes}
  \end{threeparttable}
\end{center}
\end{table*}

\subsection{Histogram of Relative Orientation}

We adopted the method of \cite{2013ApJ...774..128S} to calculate the HRO. The density gradient is calculated by convolving the data cube with a $3\times 3$ Gaussian derivative kernel in each of the three directions $x$, $y$, and $z$ (see Eq.~(6) in \cite{2013ApJ...774..128S} for the example of Gaussian derivative kernels in 2D). The angle between the density gradient $\nabla n$ and magnetic field $\mathbf{B}$ at each point in the 3D simulation cube can then be calculated as\footnote{Note that we adopted a formula for the relative angle that is different from that of \cite{2013ApJ...774..128S}.}
\begin{equation}
\phi = \arccos\left(\frac{\nabla n\cdot \mathbf{B}}{|\nabla n|~|\mathbf{B}|}\right).
\end{equation}
In 2D projected map with integrated column density $N$ and synthetic polarized emission $\boldsymbol{p}$, this becomes
\begin{equation}
\phi = \arccos\left(\frac{\nabla N\cdot \boldsymbol{p}}{|\nabla N|~|\boldsymbol{p}|}\right),
\end{equation}
where the pseudo-vector $\boldsymbol{p}$ is defined as
\begin{equation}
\boldsymbol{p} = \left(p\sin\chi\right) \hat{\mathbf{x}} + \left(p\cos\chi\right)\hat{\mathbf{y}}
\label{pvec}
\end{equation}
(see Section~\ref{sec:pol} for derivations of polarization fraction $p$ and angle $\chi$).
The HROs are therefore defined as the histograms of $\cos\phi$ (range from $-1$ to $1$) for 3D data cube, or the histogram of $\phi$ (range from $0^\circ$ to $180^\circ$) for 2D projected map \citep[cf.~][]{2013ApJ...774..128S}.

The HRO shape can provide important information of the magnetic field direction relative to the gas structure; a concave HRO means that the magnetic field is preferentially aligned perpendicular ($\phi \sim 90^\circ$ or $\cos\phi\sim 0$) to the density gradient, or parallel to the gas structure. On the other hand, a convex HRO indicates the magnetic field is perpendicular to the elongation direction of gas clumps. Determining what causes the HRO to change shape could therefore help us understand the physical processes involved in molecular cloud evolution.

Similar to \cite{2013ApJ...774..128S}, we investigate the relative orientation under different physical conditions using segmentation by density, or column density in 2D. However, we do not require our segmentation to contain equal numbers of points in each bin, as our goal is to study how the density impacts the shape of HROs. We aim to find the transition density where the HRO changes shape and the physical interpretation of this density. 
We discuss the details in the following sections. Also note that we chose to normalize the HROs so that the sum of the heights of all bars in each distribution is unity, which makes our study comparable with \cite{2013ApJ...774..128S} because the number of points in each segmentation is factored out by this normalization method.

\subsection{Derivation of Polarized Emission}
\label{sec:pol}

\begin{figure}
\begin{center}
\includegraphics[width=\columnwidth]{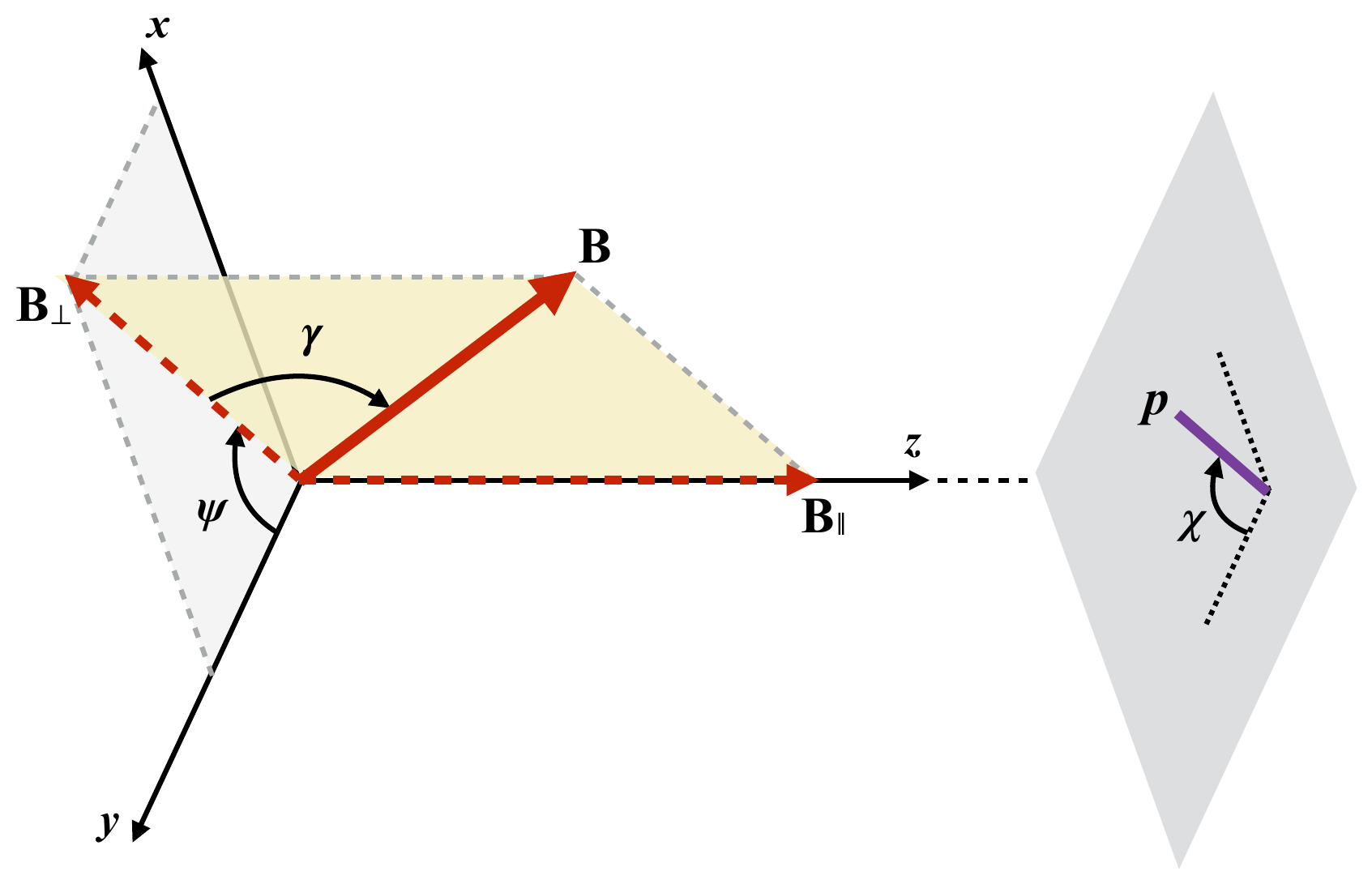}
\caption{Definition of angles. The plane of sky is in the $x$-$y$ plane, and the observer is viewing along the $-z$ direction. The inclination angle $\gamma$ is defined as the angle between the magnetic field $\mathbf{B}$ and the plane of sky, and the position angle $\psi$ is defined as the angle between the plane-of-sky magnetic field component $\mathbf{B}_\perp$ and the $y$ direction. The integrated polarization angle $\chi$ has the same convention as $\psi$. }
\label{angDef}
\end{center}
\end{figure}

The procedure we use to obtain the Stokes parameters is based on previous work \citep{1985ApJ...290..211L,2000ApJ...544..830F,2012ApJ...761...40K} wherein the relative Stokes parameters are calculated, which are functions only of the density $n$ and magnetic field structure:

\begin{subequations}
\begin{align}
q &= \int n \cos 2\psi \cos^2 \gamma ~ds, \\
u &= \int n \sin 2\psi \cos^2 \gamma ~ds.
\end{align}
\label{poldef}
\end{subequations}

\noindent Here, $\gamma$ is the inclination angle of the local magnetic field relative to the plane of the sky, and $\psi$ is the position angle of the magnetic field on the plane of the sky (see Figure~\ref{angDef}). Using a Cartesian coordinate system where the $y$-axis corresponds to North in galactic coordinates and where the $z$-axis is parallel to the line of sight, we can represent the relative Stokes parameters in a more convenient form:

\begin{subequations}
\begin{align}
q &= \int n \frac{B_y^2 - B_x^2}{B^2} ~ds,\\
u &= \int n \frac{2B_xB_y}{B^2} ~ds. 
\end{align}
\label{poldef2}
\end{subequations}

Under assumptions of perfect grain alignment and pure emission from dust grains,\footnote{Note that, though the assumption of perfect grain alignment is a simplification of real situation \citep[see e.g.][]{1998ApJ...499L..93A}, it is commonly adopted \cite[e.g.][]{2000ApJ...544..830F,2013ApJ...774..128S} and has some support from observations. 
BLASTPol \citep{2016ApJ...824..134F} reveals a flat spectrum of polarization fraction between $250-500~\mu$m for Vela C molecular cloud \citep{2016ApJ...824...84G}, suggesting the radiative environment of the gas is not playing a huge role in this region. From previously published starlight polarization data in molecular clouds, \cite{2016arXiv160509371S} also concluded that the magnetic field structure is more important in controlling polarization fraction and dispersion than grain alignment efficiency.
However, we caution the readers that the grain alignment efficiency does decrease in denser regions \citep[see e.g.][]{2015ARA&A..53..501A}, and the assumption of perfect grain alignment we adopted here is a rough approximation.
}
 it is possible to extract the relative Stokes parameters from observations.
The polarization fraction is therefore given by

\begin{equation}
p = p_0 \frac{\sqrt{q^2 + u^2}}{N - p_0 N_2},
\end{equation}

\noindent where 
$p_0$ is an intrinsic normalization factor determined by the effective polarization cross section of dust grains ($p_0 \sim 0.1$ from observations; see Section~\ref{sec:intro}),
$N$ is the column density, and $N_2$ is a correction to the column density for inclination:\footnote{Note that we did not adopt the revised version of $N_2$ claimed by \cite{2015A&A...576A.105P}. Though the difference is negligible because of the small value of $p_0$, we refer the readers to their Equation~(5) for a corrected form of $N_2$.}

\begin{equation}
N_2 = \int n\left(\frac{\cos^2 \gamma}{2} - \frac{1}{3}\right) ~ds.
\end{equation}

\noindent The inferred polarization angle on the plane of sky is given using the four-quadrant inverse tangent (which returns value in the range of $[-\pi,\pi]$ based on the signs of the two inputs),

\begin{equation}
\chi = \frac{1}{2} \text{arctan2}(u,q).
\end{equation}
The polarization angle $\chi$ is measured clockwise from North, with the same convention as the magnetic field position angle $\psi$ (see Figure~\ref{angDef}).

To focus our analysis on the compressed region, we only consider simulation cells with density $10^4 < n < 10^7$~cm$^{-3}$ in our column density and polarization fraction calculations, unless otherwise noted (e.g.~Figure~\ref{dcutmap}). Since most polarization observations of molecular clouds trace only dense gas, and the diffuse background has been removed from the observed thermal emission \citep[e.g.][]{2006ApJ...648..340L,2016ApJ...824..134F}, applying a density limit makes the simulated polarization maps more comparable to observations \citep[see e.g.][]{2008ApJ...679..537F}.

\section{HRO and The Threshold Density in field$-$density Relation}
\label{sec:BvsD}

\subsection{Choice of Segmentation for HROs}
\label{sec:ncut}

\begin{figure*}
\begin{center}
\includegraphics[width=\textwidth]{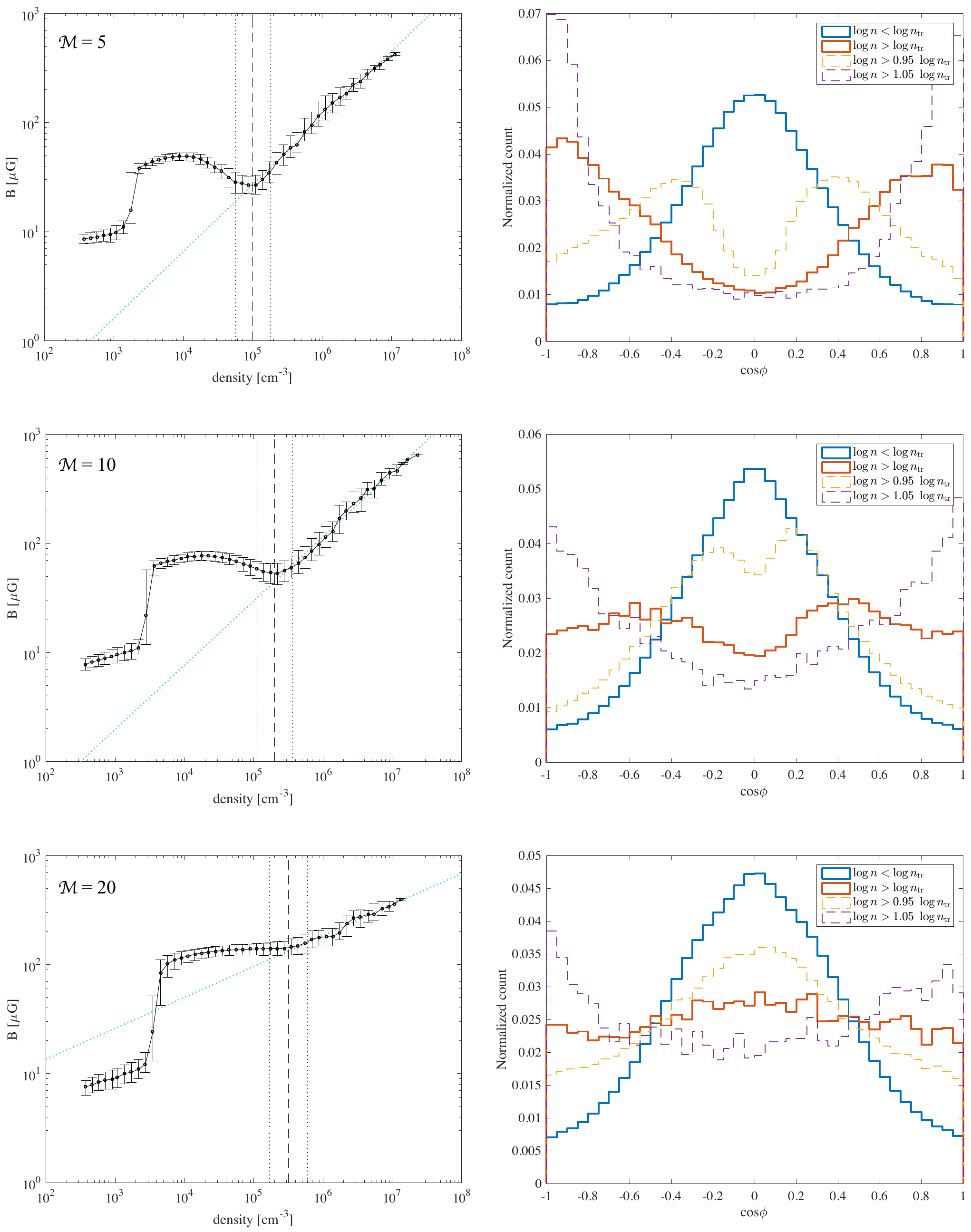}
\caption{The median and $\pm 25\%$ values of the magnetic field strength as functions of binned gas number densities ({\it left}), and the HROs ({\it right}) for models with inflow Mach numbers ${\cal M} = 5$ ({\it top}), $10$ ({\it middle}), and $20$ ({\it bottom}). Binwidth is $\Delta\left(\log n\right) = 0.1$. HROs with $n > n_\mathrm{tr}$ (dashed lines in $B-n$ plots) are distinguishable in shape from those with $n < n_\mathrm{tr}$. Dotted vertical lines in $B-n$ plots correspond to $\log n = 0.95 \log n_\mathrm{tr}$ and $\log n = 1.05 \log n_\mathrm{tr}$, respectively. The blue dotted lines ({\it left panel}) are the best linear fit to the $B-n$ relation in the self-gravitating regime.}
\label{BvsD_Ma}
\end{center}
\end{figure*}


In the ideal MHD limit, gas is strictly coupled to magnetic field. When gas is compressed by a supersonic, super-Alfv{\'e}nic shock, the magnetic field component perpendicular to the shock flow is amplified as well (see \cite{2012ApJ...744..124C} and \hyperlink{CO14}{CO14} for discussions about oblique MHD shocks). The post-shock region is therefore strongly magnetized.
However, through anisotropic condensation along the magnetic field lines, local dense clumps are able to collect materials without adding more magnetic support. As a result, the magnetic field strength is largely unrelated to the gas density. This process is shown to be dominant in strongly-magnetized shock-compressed regions, until the dense structures become dense enough to collapse gravitationally (\hyperlink{CO15}{CO15}). Self-gravitating gas would therefore show positive correlation between density and magnetic field.
Similar behavior of the field$-$density relation has been reported in numerical studies of molecular cloud formation \citep{2008A&A...486L..43H,2009ApJ...695..248H}.

\begin{figure*}
\begin{center}
\includegraphics[width=\textwidth]{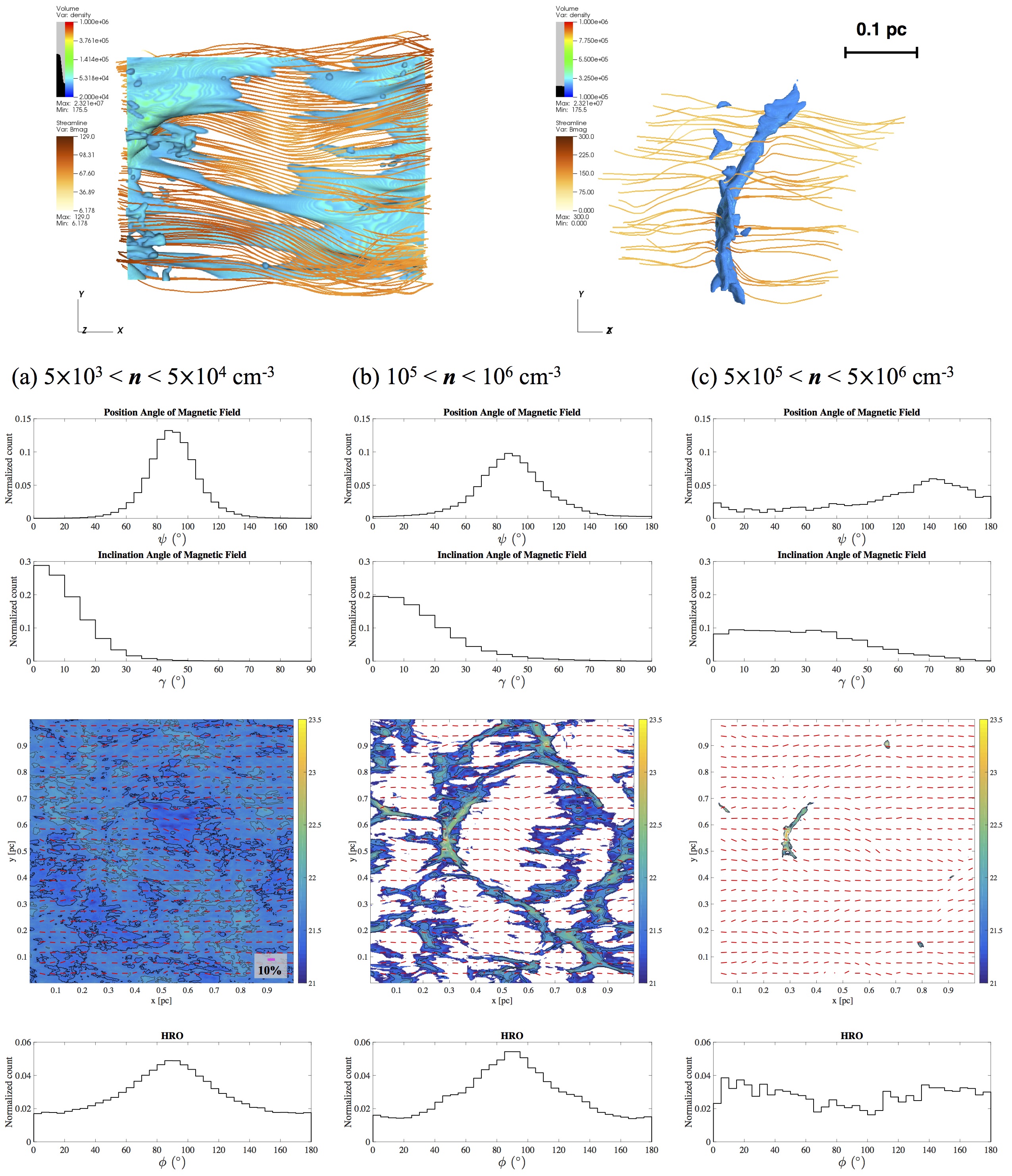}
\caption{Magnetic field directions relative to gas structures at different density ranges in model M10. {\it Top panel:} 3D illustrations of magnetic field ({\it orange lines}) and density ({\it isosurfaces}) in two sub-regions.
At lower densities ({\it left}), the filamentary network is elongated along the magnetic field lines, similar to the striations found in observations \citep[see review in][]{2014prpl.conf...27A}. At higher densities ({\it right}), the gas structure aligns preferably perpendicular to the magnetic field; also, the field lines are notably bent near the dense clump. Animations of these 3D illustrations, panning about the center to better view the full structure, are available on the online journal. 
{\it Second and third rows:} histograms of the position ({\it second row}) and inclination angles ({\it third row}) of the magnetic field, from simulation cells within density ranges $5\times 10^3 < n < 5\times 10^4$~cm$^{-3}$ ({\it left}), $10^5 < n < 10^6$~cm$^{-3}$ ({\it middle}), and $5\times 10^5 < n < 5\times 10^6$~cm$^{-3}$ ({\it right}). 
The transition density in this particular simulation run is $n_\mathrm{tr}  \approx 2\times 10^5$~cm$^{-3}$ (see Table~\ref{sumtable}).
The flattened distributions of both $\psi$ and $\gamma$ at higher densities suggest that the magnetic field is indeed bent and tangled, showing quantitatively what is suggested by the 3D illustrations above.
{\it Fourth and fifth rows:} projected column density ({\it colormap}) with derived polarization ({\it pink segments; fourth row}) and the resulting HROs ({\it bottom row}), at corresponding density ranges. The change of HRO shapes in 2D projected maps can be directly linked to the gas density traced in 3D space,
and observations of molecular lines to trace gas density could, in principle, determine $n_\mathrm{tr}$ in conjunction with HRO techniques.
Note that the simulated polarization is integrated using the whole post-shock region without the corresponding density limit to be comparable with polarized continuum emission from dust grains in observations. }
\label{dcutmap}
\end{center}
\end{figure*}

For simulations with different shock Mach numbers in \hyperlink{CO15}{CO15} at the time when the most evolved core starts to collapse, we bin the dataset by the log of number density $n$ with bin width $0.1$, and calculate the median value of the magnetic field $B$ in each bin. The results are shown in Figure~\ref{BvsD_Ma} ({\it left panel}), where the error bars represent the first and third quartiles (i.e.~the $\pm 25\%$ values of the distribution). The pre-shock and post-shock regions can be easily distinguished by the sharp jump in magnetic field strength near $n\sim 10^3$~cm$^{-3}$. Between $n\sim 10^3-10^5$~cm$^{-3}$, the magnetic field strength is roughly unchanged with increasing number density, a feature predicted by the anisotropic condensation model described in \hyperlink{CO15}{CO15}.\footnote{Note that the reconnection diffusion process might also play an important role for keeping magnetic field relatively constant at low gas density regime \citep[e.g.][]{2012ApJ...757..154L,2015MNRAS.452.2500L}. }
This also leads to a relatively homogeneous magnetic field direction in the post-shock layer; Figure~\ref{dcutmap} ({\it top left}) demonstrates the well-ordered magnetic field in 3D at density $n\sim 10^4$~cm$^{-3}$ (an animation of changing viewing angles is available on the online journal).

A more important transition happens at $n \sim 10^5$~cm$^{-3}$, where the magnetic field becomes strongly correlated with the number density
(roughly as a power-law scaling; see Section~\ref{sec:bn}).
This suggests that the magnetic field is compressed with the gas in this regime, and the 3D visualization in Figure~\ref{dcutmap} ({\it top right}) illustrates the twisted magnetic field lines around dense clumps (an animation of changing viewing angles is available on the online journal).
In fact, we found that magnetic field morphology changes with density in the simulations.
Figure~\ref{dcutmap} ({\it second and third rows}) shows the statistics of magnetic field directions at densities below ({\it left}), around ({\it middle}), and above ({\it right}) this transition value.\footnote{Also shown in Figure~\ref{dcutmap} is the HRO of polarization fraction and column density gradient ({\it bottom row}), which we will discuss in detail in Section~\ref{sec:pvsN}.}
As the gas density increases, both the distributions of position ($\psi$) and inclination angle ($\gamma$; see Figure~\ref{angDef} for definition) of magnetic field spread to wider ranges, and thus the uniformity of magnetic field decreases. The change is dramatic for densities larger than the transition value of the $B-n$ curve; the plane-of-sky magnetic field loses the preferred direction at $\psi\approx 90^\circ$, and the total magnetic field can no longer be considered roughly parallel to the plane of sky with $\gamma \sim 0^\circ$.

We therefore define the transition density at which the magnetic field and gas density become highly correlated,
$n_\mathrm{tr}$, 
as the boundary value of the segmentation when drawing HROs. The value of $n_\mathrm{tr}$ for each model is listed in Table~\ref{sumtable}. In the right panel of Figure~\ref{BvsD_Ma} we plotted HROs corresponding to density segmentations with $n < n_\mathrm{tr}$ ({\it blue lines}) and $n > n_\mathrm{tr}$ ({\it red lines}).
The HRO shape clearly changes from concave to convex when density increases from $n<n_\mathrm{tr}$ to $n>n_\mathrm{tr}$. 
To demonstrate that the onset of HRO shape change is definitely correlated with $n_\mathrm{tr}$, we also included in Figure~\ref{BvsD_Ma} the $95\%$ and $105\%$ values of $\log n_\mathrm{tr}$ and the corresponding HROs. There is significant change in HRO shape even with small deviations from $n_\mathrm{tr}$. 
This is further evidence that $n_\mathrm{tr}$ marks a transition between two different HRO shapes. 

\begin{figure}
\begin{center}
\includegraphics[width=\columnwidth]{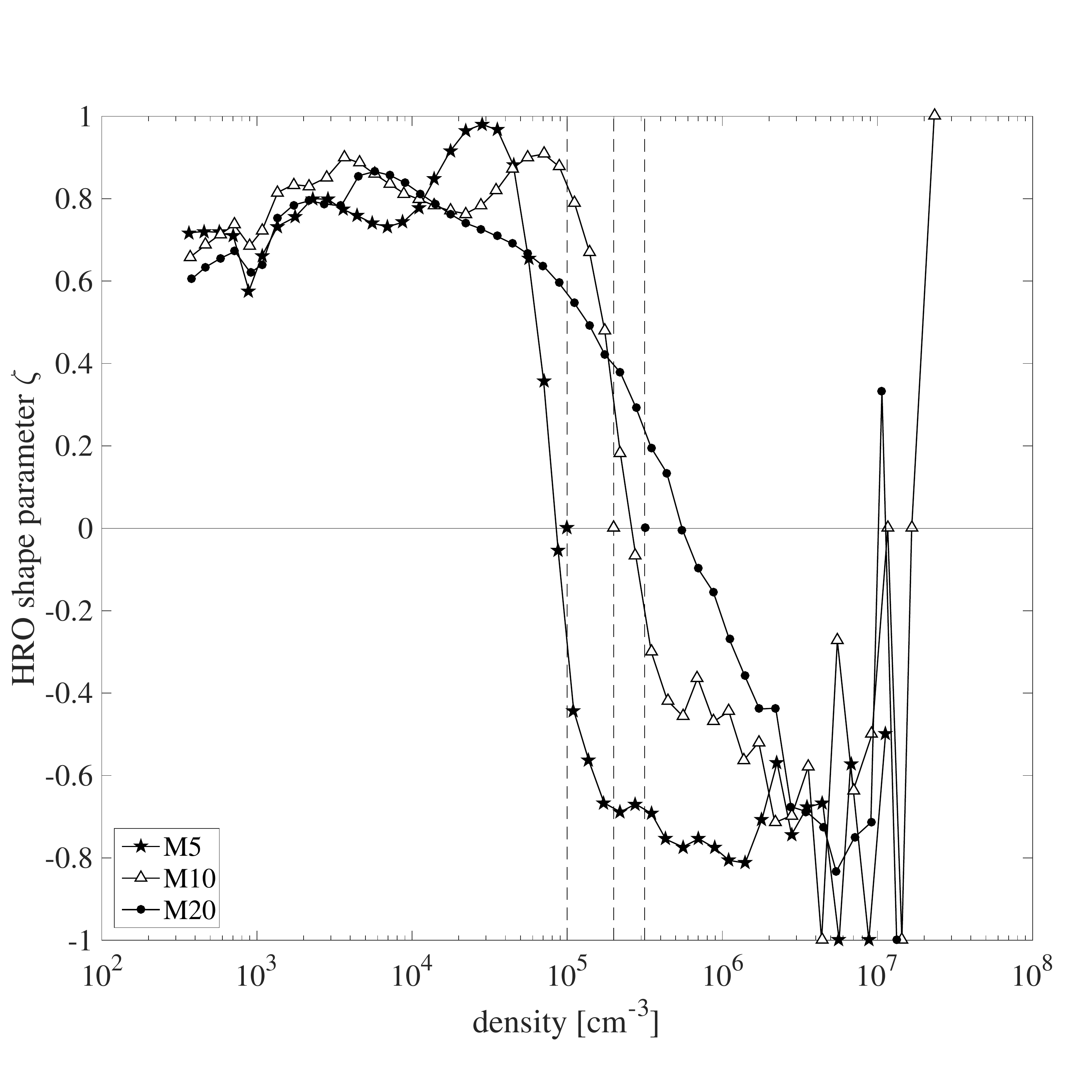}
\caption{HRO shape parameter $\zeta$ in volume density segmentations for models M5 ({\it stars}), M10 ({\it triangles}), and M20 ({\it filled circles}); $\zeta \approx 0$ indicates the transition of HRO shape from concave to convex. Vertical dashed lines represent $n_\mathrm{tr}$ defined from Figure~\ref{BvsD_Ma} and are labeled by corresponding marks for each model.}
\label{Bn_zeta}
\end{center}
\end{figure}

In addition, we calculated the HRO shape parameter $\zeta$ as defined in \cite{2016A&A...586A.138P} (see their Equation~(4)):\footnote{This is an updated definition of that from the original HRO paper \cite{2013ApJ...774..128S}.}
\begin{equation}
\zeta \equiv \frac{A_c - A_e}{A_c + A_e},
\label{zetaDef}
\end{equation}
where $A_c$ and $A_e$ represent the $25\%$ areas under the HRO curve at the center ($-0.25<\cos\phi<0.25$, or $67.5^\circ<\phi<112.5^\circ$ in 2D) and on the edge ($-1.00<\cos\phi<-0.75$ and $0.75<\cos\phi<1.00$, or $0^\circ < \phi < 22.5^\circ$ and $157.5^\circ < \phi < 180^\circ$ in 2D), respectively. 
A concave HRO would have $\zeta > 0$, and a convex HRO, $\zeta < 0$. Thus the parameter $\zeta$ provides a way to characterize HRO shape in a single parameter.
The values of $\zeta$ as functions of number density for all three models are plotted in Figure~\ref{Bn_zeta}, as well as the transition densities $n_\mathrm{tr}$ defined from $B-n$ plots in Figure~\ref{BvsD_Ma}. The significant change of $\zeta$ from positive to negative around $n_\mathrm{tr}$ is striking, and the threshold density of the $B-n$ relation predicts the transition of HRO shape very well.
We discuss the physical reason for the connection between HRO shape and $n_\mathrm{tr}$ in the following section.

\subsection{HRO Shape and Gas Self-gravity}
\label{sec:gravity}

\begin{figure*}
\begin{center}
\includegraphics[width=0.95\textwidth]{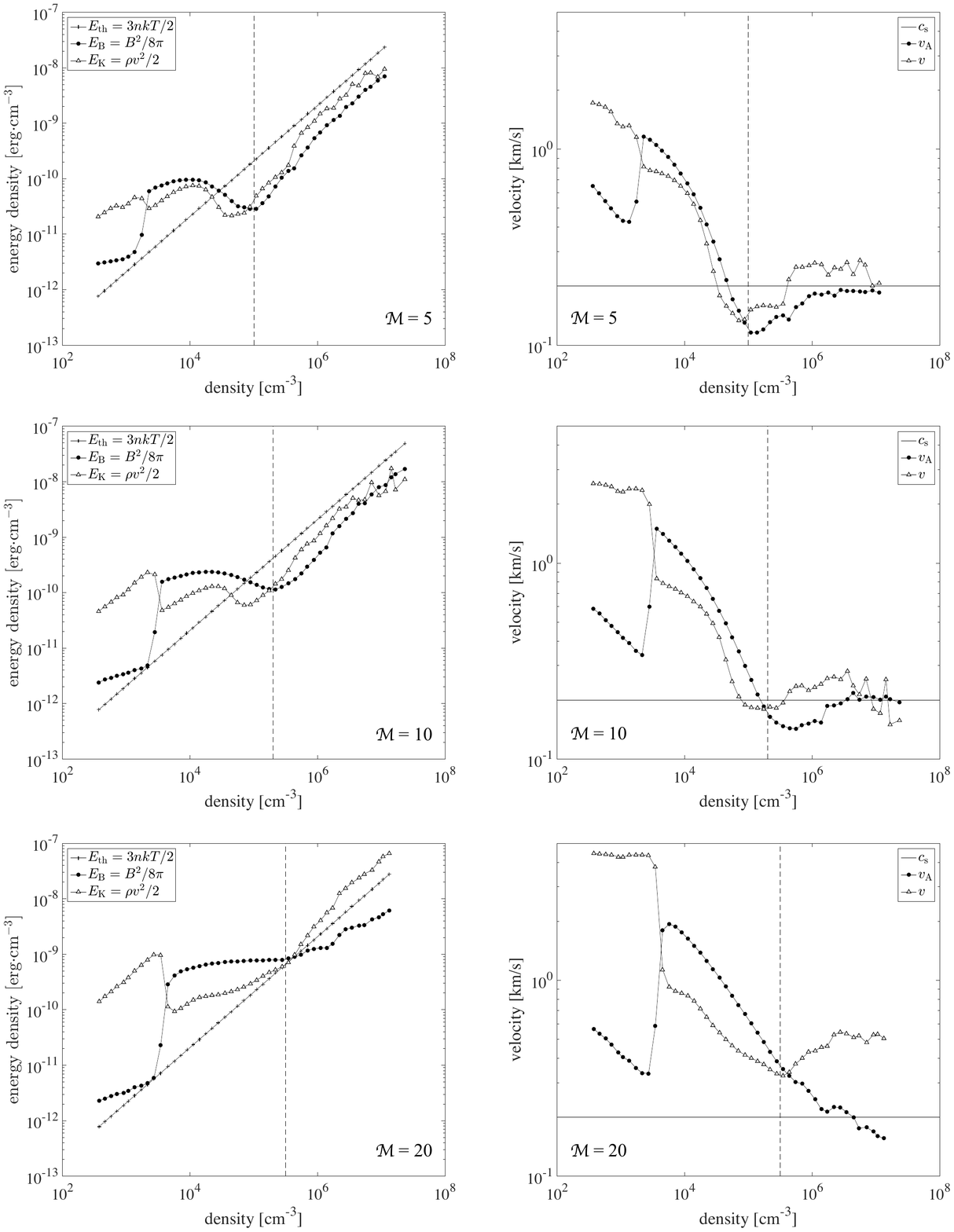}
\caption{The median values of thermal ({\it plus signs}), magnetic ({\it filled circles}), and kinetic ({\it triangles}) energy densities ({\it left panel}) and corresponding velocity magnitudes (sound speed $c_\mathrm{s}$, Alfv\'en speed $v_\mathrm{A}$, and gas velocity $v$; {\it right panel}) in each density bin, for models M5 ({\it top}), M10 ({\it middle}), and M20 ({\it bottom}). 
The dashed line marks the boundary value for HRO segmentation ($n_\mathrm{tr}$) in each model.
The gas is supersonic and dominated by the kinetic energy from large-scale turbulence in the pre-shock region, but becomes strongly confined by magnetic pressure downstream.
The kinetic energy is therefore relatively unchanged (gas velocity decreases with density) in the post-shock region, until density reaches the threshold density $n_\mathrm{tr}$. 
The increase in kinetic energy prior to $n = n_\mathrm{tr}$ is apparently induced by the self-gravity of the gas, and $n_\mathrm{tr}$ marks the onset of gravity-induced acceleration in the gas; this is also the transition from sub-Alfv\'enic to super-Alfv\'enic gas flow (i.e.~the equipartition of magnetic and kinetic energies).
}
\label{energy}
\end{center}
\end{figure*}

As mentioned in previous sections, the self-gravity of magnetized gas can induce gravitational collapse that compresses both the gas and the magnetic field. The threshold density $n_\mathrm{tr}$ $-$ where the field strength becomes a power law in density, or where the HRO shape changes from concave to convex $-$ therefore corresponds to the density at which the gas becomes self-gravitating. 
However, the gravitational energy cannot be determined uniquely, as it depends on where the reference point of the gravitational potential is chosen. Nonetheless, the effect of gravity can be inferred through its dynamical influence on the gas. 
We can therefore use the increases in gas kinetic energy and velocity magnitudes to gauge the significance of self-gravity. 

Figure~\ref{energy} compares the thermal ($E_\mathrm{th}$), magnetic ($E_\mathrm{B}$), and kinetic ($E_\mathrm{K}$) energy density of the gas ({\it left panel}), as well as the corresponding velocity magnitudes: sound speed $c_\mathrm{s}$, Alfv\'en speed $v_\mathrm{A} \equiv B/\sqrt{4\pi\rho}$, and gas velocity $v$ ({\it right panel}), as functions of density. It shows that, the kinetic energy first drops from the initial values at the shock front, making the post-shock region dominated by the magnetic pressure so that anisotropic gas flow along the magnetic field is preferred. The gas velocity therefore decreases with density in this regime because of the lack of energy input, resulting in a relatively unchanged kinetic energy.
At a density somewhat below $n_\mathrm{tr}$ ({\it dashed lines}), $E_\mathrm{K}$ starts to rise (or, in the case of model M20, the slope of $E_\mathrm{K}$ becomes steeper); then the gas transitions from $E_\mathrm{K} < E_\mathrm{B}$ to $E_\mathrm{K} > E_\mathrm{B}$ close to $n = n_\mathrm{tr}$ for all three cases. 
More importantly, $n_\mathrm{tr}$ clearly marks the beginning of gravity-induced acceleration in terms of increasing gas velocity with density. This happens right at the intersection of $v$ and $v_\mathrm{A}$, i.e.~at the transition for the gas to become super-Alfv\'enic ($v > v_\mathrm{A}$) from sub-Alfv\'enic ($v < v_\mathrm{A}$).

\begin{figure}
\begin{center}
\includegraphics[width=\columnwidth]{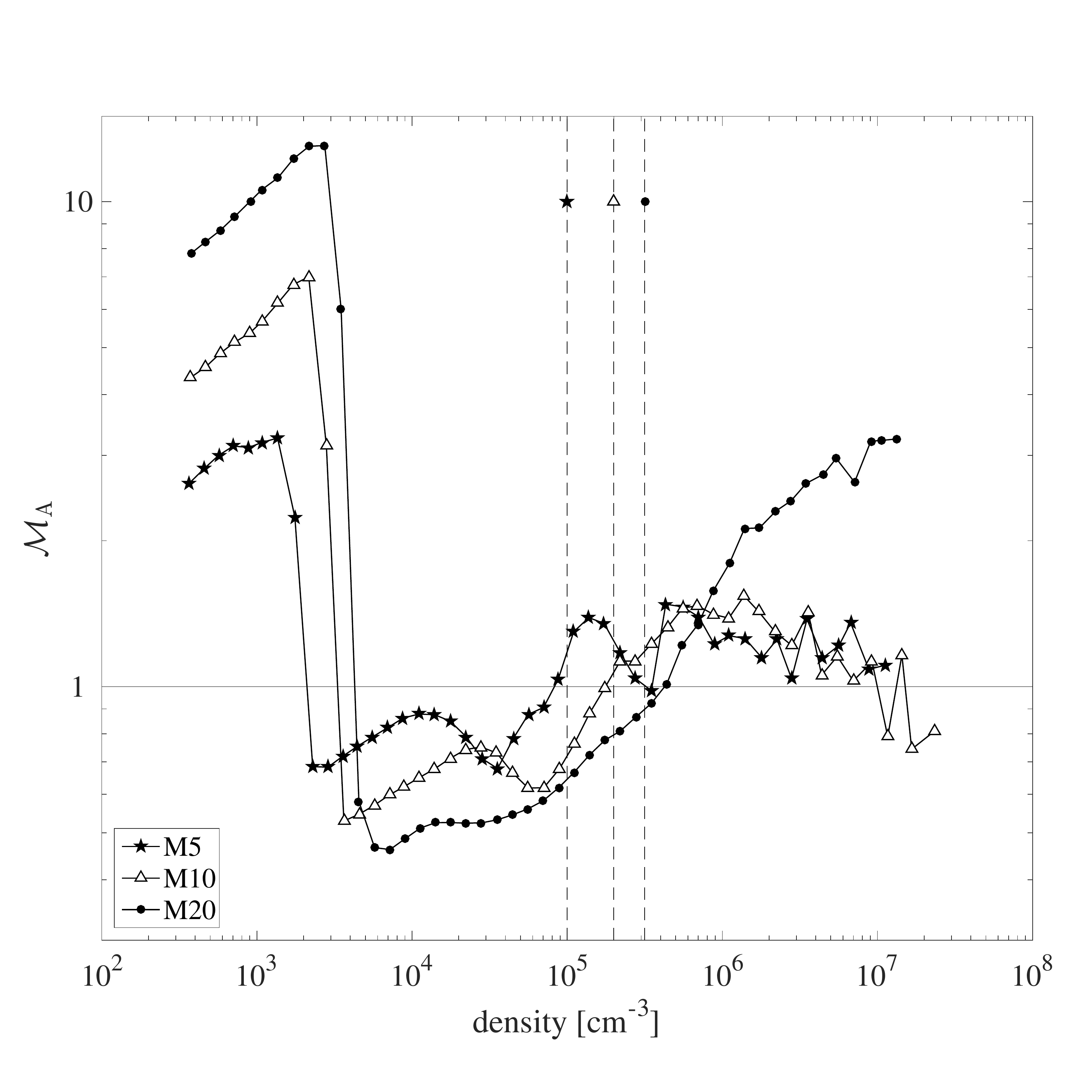}
\caption{Alfv{\'e}n Mach number ${\cal M}_\mathrm{A}$ as a function of density for models M5 ({\it stars}), M10 ({\it triangles}), and M20 ({\it filled circles}). Also plotted are the boundary values for HRO segmentation, $n_\mathrm{tr}$, measured from the $B-n$ relation in Figure~\ref{BvsD_Ma} ({\it dashed lines with corresponding marker}). Note that model M20 is the only one that stays super-Alfv{\'e}nic even at density scales of dense cores ($n \gtrsim 10^6$~cm$^{-3}$).}
\label{MA}
\end{center}
\end{figure}

We can also directly compare $v$ and $v_\mathrm{A}$, or the relative importance of kinetic energy and magnetic energy, by measuring the Alfv{\'e}n Mach number:
\begin{equation}
{\cal M}_\mathrm{A} \equiv \frac{v}{v_\mathrm{A}} = \frac{v}{B/\sqrt{4\pi\rho}}.
\label{MAdef}
\end{equation}
Figure~\ref{MA} shows the behavior of ${\cal M}_\mathrm{A}$ in each model as a function of gas density. The plot demonstrates a clear transition from the super-Alfv{\'e}nic (${\cal M}_\mathrm{A} > 1$) pre-shock flow to the sub-Alfv{\'e}nic (${\cal M}_\mathrm{A} < 1$) post-shock layer, within which overdense regions gradually acquire kinetic energy by falling into local gravitational potential wells, get denser, and become super-Alfv{\'e}nic at $n \approx n_\mathrm{tr}$ ({\it dashed lines}).\footnote{Figure~\ref{MA} also illustrates the difficulty with using ${\cal M}_\mathrm{A}$ as the segmentation criterion of HRO in simulations like ours where the behavior of ${\cal M}_\mathrm{A}$ in different regimes (pre-shock, post-shock; non-self-gravitating, self-gravitating) is non-monotonic, with voxels in different evolutionary stages potentially sharing the same values of ${\cal M}_\mathrm{A}$.}
Thus $n_\mathrm{tr}$ represents a characteristic value for gravity-induced Alfv\'enic transition in a strongly magnetized cloud: gas dynamics at $n < n_\mathrm{tr}$ is constrained by magnetic pressure, while at $n > n_\mathrm{tr}$ the gas kinetic energy (augmented by gravity) dominates.
Note that there may be other reasons for the Alfv{\'e}n Mach number ${\cal M}_\mathrm{A}$ to increase with increasing density. For example, \cite{2009ApJ...693..250B} found a similar trend in non-self-gravitating, globally sub-Alfv{\'e}nic clouds, most likely because the mass density increases faster than the magnetic energy density for condensation along field lines, causing the Alfv{\'e}n speed $v_\mathrm{A}$ to drop with density. This drop is also seen in our model M20 (see Figure~\ref{energy}). In this case, the drop in Alfv{\'e}n speed $v_\mathrm{A}$ contributes to the increase in ${\cal M}_\mathrm{A}$. For the other two cases (M5 and M10), $v_\mathrm{A}$ increases with density beyond the transition density $n_\mathrm{tr}$, which means that the increase in ${\cal M}_\mathrm{A}$ must come entirely from the increase in flow speed. In other words, the flow speed is increased so much that the Alfv{\'e}n Mach number increases with density despite the increase in Alfv{\'e}n speed. Gravitational contraction naturally explains both the increase in flow speed and the increase in Alfv{\'e}n speed because it causes condensation across field lines, which tends to increase the magnetic energy density faster than the mass density.

These results provide a natural explanation for why the magnetic field starts to correlate positively with density at $n > n_\mathrm{tr}$: the magnetic pressure in this density regime is no longer strong enough to resist  condensation across the field lines. On the other hand, sub-Alfv\'enic, anisotropic flow along the magnetic field lines is favored at $n < n_\mathrm{tr}$, and it is difficult for subregions to contract perpendicular to the field lines in this regime. As a result, gas structures will gradually evolve to become elongated roughly perpendicular to the magnetic field direction, especially under the influence of self-gravity (see \hyperlink{CO15}{CO15}). This morphology is reflected in the convex HRO at $n > n_\mathrm{tr}$ when cross-field line contraction is just able to begin.
At $n < n_\mathrm{tr}$, 
the HRO remains concave because it continues to be
dominated by the low-density striations that are preferentially aligned with the field lines (see Figure~\ref{dcutmap}, {\it top left}).\footnote{Network of small sub-filaments, or striations, is a common feature of strongly magnetized cloud, both observationally (\citealt{2008ApJ...680..428G,2011ApJ...734...63S,2012A&A...543L...3H,2013A&A...550A..38P}; also see review in \citealt{2014prpl.conf...27A}) and theoretically (\citealt{1995ApJ...438..763G,2003ApJ...590..858V,2008ApJ...680..420H}; \hyperlink{CO14}{CO14}; \hyperlink{CO15}{CO15}), though its formation mechanism is still under debate.}
The transformation of HRO shape is therefore directly related to the gravity-driven Alfv\'enic transition at the threshold density $n_\mathrm{tr}$.

\subsection{The Field$-$density Relation}
\label{sec:bn}

Figure~\ref{BvsD_Ma} ({\it left panel}) indicates that the high-density gas is self-gravitating and thus the magnetic flux is compressed along with the gas, resulting in a strong correlation between magnetic field strength $B$ and gas number density $n$ that can be fitted by a power-law scaling $B\propto n^\kappa$.
Numerous studies, both observational and theoretical, have been conducted to determine the value of $\kappa$ in this so-called field$-$density relation \citep[e.g.][]{1993ApJ...415..680F,1999ApJ...520..706C,1999ApJ...526..279P,2001ApJ...546..980O,2010ApJ...725..466C,2012ApJ...755..130M,2014MNRAS.437...77B,2015Natur.520..518L, 2015MNRAS.452.2500L, 2015MNRAS.451.4384T}. 

For simulation models discussed in the previous section, we calculated the best-fit value of $\kappa$ for $n > n_\mathrm{tr}$ regime, and overplotted the derived relation $B\propto n^\kappa$ in Figure~\ref{BvsD_Ma} ({\it left panel}). The fitted $\kappa$ values are listed in Table~\ref{sumtable}. Our $\kappa$ value varies from $\sim 0.3$ to $0.6$ between models, which is consistent with the observed $0.47-0.65$ value in dense gas \citep{1999ApJ...520..706C,2010ApJ...725..466C}. Our result suggests a strong dependence of $\kappa$ on the physical conditions of the simulations; models with stronger inflows (which create post-shock regions with a lower plasma $\beta$ value and higher velocity perturbation; see Table~\ref{sumtable}) have a weaker $B-n$ dependence. 

The dependence of $\kappa$ on turbulent level that we found is qualitatively similar to the numerical results reported in \cite{2001ApJ...546..980O}. Though the range of density covered in \cite{2001ApJ...546..980O} is different from this study, the positive correlation between magnetic field and density at high densities is clear. Also, for the strongly magnetized regime (their $\beta_0<1$ models), a weaker supersonic turbulence tends to produce a stronger $B-n$ correlation (see their Figures 18 and 19), similar to what we found. In addition, a recent study by \cite{2015MNRAS.452.2500L} adopted ${\cal M} = 10$ as the initial turbulent strength and found $\kappa \approx 0.70$ and $0.57$ for models with $\beta_0 = 0.02$ and $2.0$, respectively. The $\kappa$ value we found in our model M10 (which has $\beta_0 \approx 0.4$), $0.59$, is in good agreement with their results. This also indicates a possible trend of lower $\kappa$ values in strongly magnetized (small $\beta$) media.
To conclude, the $B-n$ scaling may not be universal, with the exponent $\kappa$ affected by the magnetic and kinematic properties of the molecular cloud.

\section{HRO and Simulated Polarization Emission}
\label{sec:pvsN}

\subsection{From $n_\mathrm{tr}$ to $N_\mathrm{tr}$}

In 3D data, there is a drastic change in the HRO shape as the gas transitions from a sub-Alfv\'enic state to a super-Aflv\'enic state through gravitational acceleration, at the threshold density $n_\mathrm{tr}$ in the magnetic field$-$density relation. However, neither the total magnetic field strength nor the volume density is easily measurable observationally. What can be measured more easily are the column density along the line of sight, from integrated thermal continuum or molecular line emission \citep[e.g.][]{1994ApJS...95..419D,2010A&A...518L..88B,2010A&A...518L.106K,2011A&A...533A..94H}, and polarized dust emission, which can provide information of the magnetic field direction projected on the plane of sky (see Section~\ref{sec:intro}).

\begin{figure*}
\begin{center}
\includegraphics[width=\textwidth]{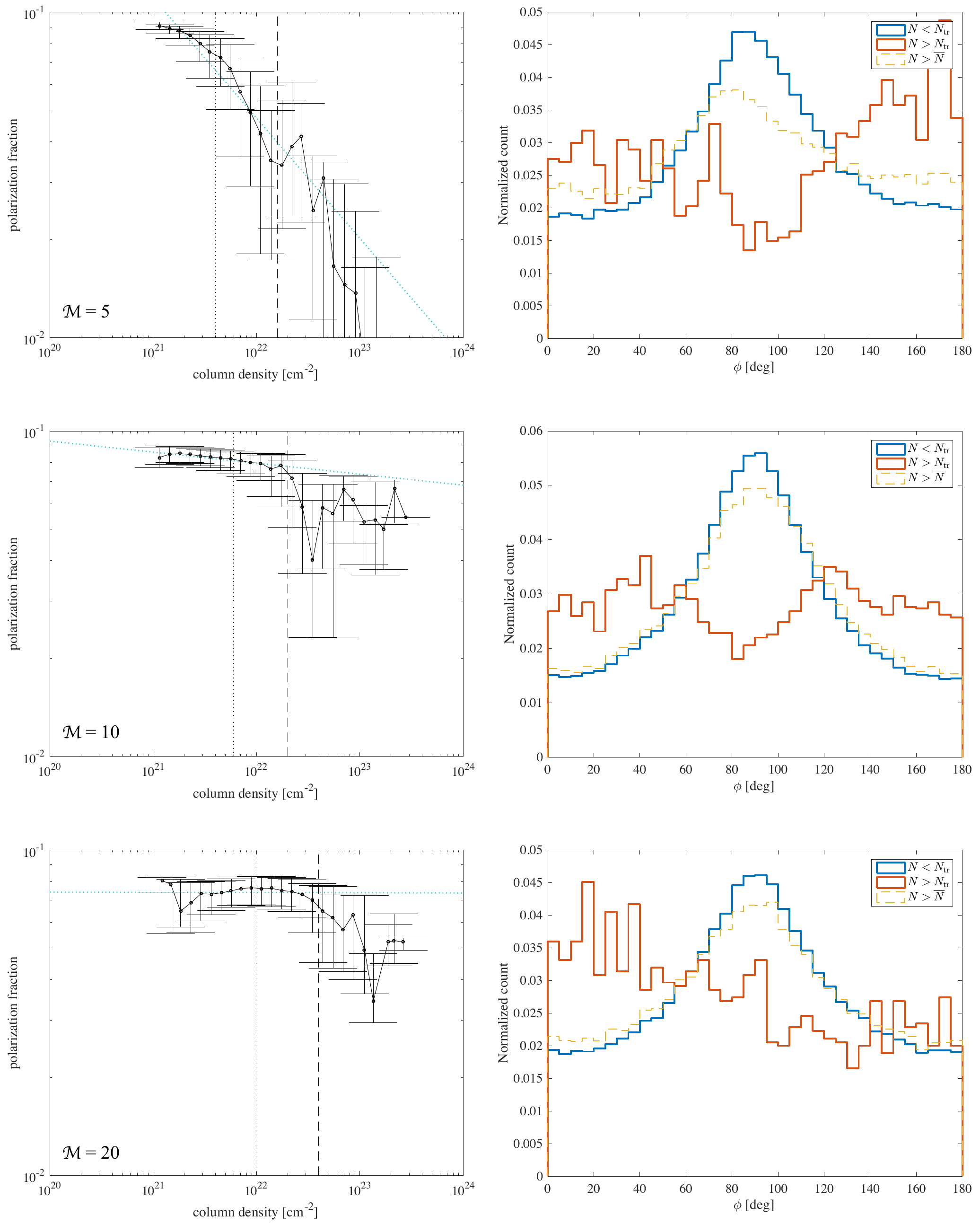}
\caption{The median and $\pm 25\%$ values of the polarization fraction as functions of binned gas column densities ({\it left}), and the HROs ({\it right}) for models with inflow Mach numbers ${\cal M} = 5$ ({\it top}), $10$ ({\it middle}), and $20$ ({\it bottom}). Dotted lines and dashed lines in $p-N$ plots correspond to the mean column density $\overline{N}$ and the transition column density $N_\mathrm{tr}$, respectively. HROs with $N > N_\mathrm{tr}$ are distinguishable in shape from those with $N < N_\mathrm{tr}$ or $N > \overline{N}$. The best-fitted $p-N$ relations in $N < N_\mathrm{tr}$ regimes are the light blue dotted lines in the left panel.}
\label{pvsN_Ma}
\end{center}
\end{figure*}

We therefore investigated the polarization fraction $p$ as a function of column density $N$ in the 2D projected map. The polarization fraction is calculated based on the method discussed in Section~\ref{sec:pol}, and we chose $p_0 = 0.1$ as the maximum polarization fraction to be comparable with observational results (see discussions and references in Section~\ref{sec:intro}).\footnote{Note that different choices of $p_0$ would simply re-normalize the simulated polarization level, and do not change any of our conclusions.} For simplicity, we chose to observe the post-shock layer face-on so that the mean magnetic field direction is on the plane of sky. The column density is thus $N(x,y) \equiv \sum_z n(x,y,z)$, where $z$ is the inflow direction (also see Figure~\ref{angDef} for the convention of coordinate system). 

We binned all pixels in the projected map by $\log N$ with bin width $0.1$, and calculated the median and $\pm 25\%$ values of the polarization fraction $p$. Figure~\ref{pvsN_Ma} ({\it left panel}) plots the $p-N$ relation from models M5, M10 and M20; in general, the polarization fraction appears to decrease with column density monotonically for lower $N$, then becomes relatively random after a sharp drop or rise.
This can be understood as increasing disorder in magnetic field structure with column density, especially when the self-gravity of the gas takes over the control. As discussed in the previous section, gravitational collapse will likely induce isotropic amplification of magnetic field. This may lead to large variations in the the inclination angle of the magnetic field with respect to the plane of sky which, together with the relatively small number of high column density cells, leads to large fluctuations in polarization fraction at the high $N$ end. 

The column density segmentation for HROs of 2D projected maps, $N_\mathrm{tr}$, is therefore defined as the column density where $p$ shows a sharp change.\footnote{Note that this transition may not be as clear-cut as in the 3D case, because the polarization fraction is an integrated quantity that provides only a rough estimate of the true onset of the gravitationally driven transition to the super-Alfv\'enic state for projected 2D structures.} This value is marked by dashed line for each model in Figure~\ref{pvsN_Ma} ({\it left panel}), and the corresponding HROs are shown in Figure~\ref{pvsN_Ma} ({\it right panel}). The change of HRO shape for column density segmentation below or above $N_\mathrm{tr}$ is evident; HROs at lower column densities are concave, then become relatively flat or convex when $N>N_\mathrm{tr}$.
Also plotted in Figure~\ref{pvsN_Ma} is the HROs with $N > \overline{N}$, the average column density of the projected map ({\it dotted lines in the left panel}). Since the transition for gas structures to become self-gravitating is a continuous process, the concave HROs with $N>\overline{N}$ simply provide a reference point that this transition must happen within column density range $\overline{N} < N < N_\mathrm{tr}$.

\begin{figure}
\begin{center}
\includegraphics[width=\columnwidth]{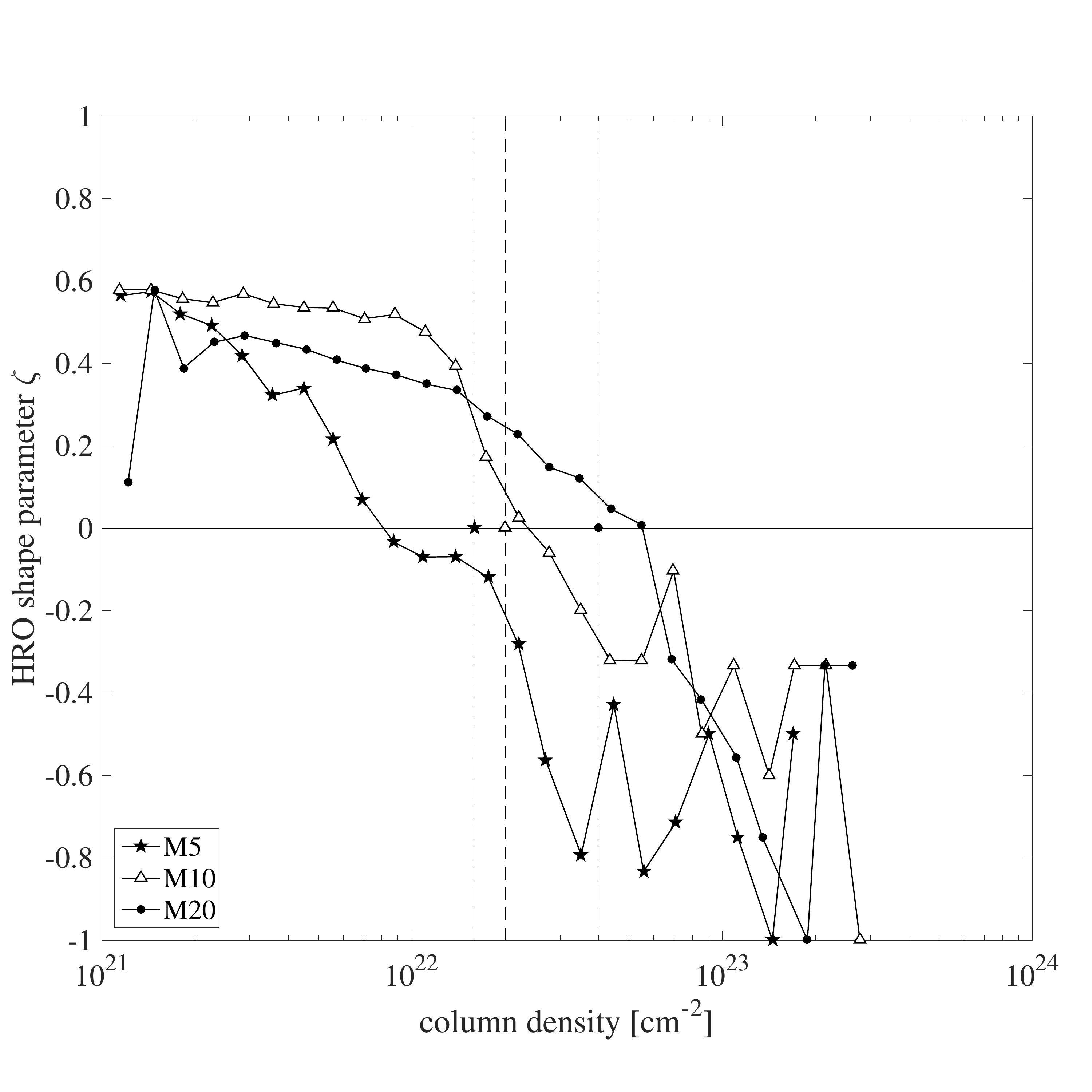}
\caption{HRO shape parameter $\zeta$ segmented by column density for M5 ({\it stars}), M10 ({\it triangles}), and M20 ({\it filled circles}); $\zeta \approx 0$ suggests the transition of HRO shape from concave to convex. Vertical dashed lines represent $N_\mathrm{tr}$ defined from Figure~\ref{pvsN_Ma} and are labeled by corresponding markers for each model.}
\label{pvsN_zeta}
\end{center}
\end{figure}

The HRO shape parameter $\zeta$, defined in Equation~(\ref{zetaDef}), for different column density segmentations is plotted in Figure~\ref{pvsN_zeta}. We again compared the $N_\mathrm{tr}$ values defined from Figure~\ref{pvsN_Ma} ({\it dashed lines with corresponding markers}) with the column densities where $\zeta = 0$. Similar to the 3D case, the threshold value identified from $p-N$ relation provides good prediction for $N (\zeta=0)$.

Note that the change of HRO shape in the 2D projected map is directly related to the gas volume density in the 3D space. Figure~\ref{dcutmap} ({\it bottom row}) shows the HRO between the projected column density gradient from simulation cells with different number densities and the integrated polarization direction. This is analogous to observing the cloud with various molecule tracers to probe gas content within different density ranges, and using continuum emission to collect polarization information (see Figure~\ref{dcutmap}, {\it fourth row}). The HRO changes from central-peaked in $n \lesssim n_\mathrm{tr}$ ({\it left two panels}) to relatively flat in $n > n_\mathrm{tr}$ ({\it right panel}), which suggests that in principle one can constrain the threshold density $n_\mathrm{tr}$ by conducting observations toward the same region with molecular lines that trace different density regimes and comparing the shape of HRO between the polarization direction from continuum and column density distributions from different density tracers. 
Furthermore, from Figure~\ref{energy} we know that $E_\mathrm{B} \approx E_\mathrm{K}$ at $n = n_\mathrm{tr}$, and $E_\mathrm{K}$ can be estimated by measuring the rms velocity from molecular line emission. Therefore, our results suggest that one can in principle obtain the value of $E_\mathrm{B}$, or the total magnetic field strength, at density $n \sim n_\mathrm{tr}$. Since $n \sim n_\mathrm{tr}$ is related to self-gravitating gas that forms prestellar/protostellar cores, the magnetic field strength in this regime is extremely important to star formation studies. We will postpone a more detailed discussion of this application to a follow-up paper (King~et al., {\it in prep.}).

\subsection{The $p-N$ Correlation}
\label{sec:pN}

Several observations have confirmed a negative correlation between the polarization fraction and column density \citep[e.g.][]{2001ApJ...562..400M,2005A&A...430..979G,2015A&A...576A.105P,2016ApJ...824..134F}, in the form $p \propto N^{-\alpha}$ or $p \propto I^{-\alpha}$ with $\alpha \sim 0.5-1.2$, where $I$ is the integrated intensity.\footnote{Note that, unlike other studies, \cite{2015A&A...576A.105P} considered a linear fit of the form $p \propto \log N$ instead of $\log p \propto \log N$. Also, while most of the studies chose to fit the entire $p-N$ or $p-I$ distribution, \cite{2015A&A...576A.105P} applied fits to the upper envelope of the distribution.} Though there are many factors that might affect the polarization degree in observations like imperfect grain alignment \citep{1999ApJ...516..834H} and dust scattering \citep{2016MNRAS.456.2794Y}, the calculated polarization fraction from our simulations is not affected by those processes. It captures the intrinsic $p-N$ correlation produced by the magnetic field geometry before it is modified by other, additional, processes.  

For the $p-N$ relation at column densities lower than the threshold value $N_\mathrm{tr}$, we fitted the curve in the form $p \propto N^{-\alpha}$ ({\it blue dotted lines} in Figure~\ref{pvsN_Ma}, {\it left panel}) and obtained the power-law index $\alpha \approx 0.0 - 0.4$ (see Table~\ref{sumtable}). The correlation between $p$ and $N$ is highly dependent on the environment. We note that our results are consistent with those discussed by \cite{2008ApJ...679..537F}, who found $\alpha \sim 0.5$ in models with supersnoic and sub-Alfv{\'e}nic flows, which is comparable to the post-shock environment in our simulations (see Table~\ref{sumtable}). Though the simulation setup in \cite{2008ApJ...679..537F} is different from what we adopted in this study, the general behavior of MHD gas dynamics is similar in the way that polarization fraction decreases with increasing randomness of the magnetic field, which is induced by compressive gas flows that also enhance column density. 

\section{Controlling Factors of Polarization Fraction}
\label{sec:pfactors}

In previous sections we have shown that the magnetic field morphology changes significantly during the gravity-induced Alfv\'enic transition, and the observed polarization level varies accordingly, which can be utilized to indicate the magnetic field strength. Here, in order to better understand the relation between magnetic field morphology and resulting polarization fraction, we provide a quantitative analysis of what causes the variation of polarization fraction in star-forming GMCs.

\subsection{Tangledness and Inclination}

From Equation~(\ref{poldef}), the polarization fraction $p$ and angle $\chi$ are affected by the inclination angle $\gamma$ of the magnetic field with respect to the plane of sky, and the dispersion of the plane-of-sky component of the magnetic field ($\mathbf{B}_\perp$) along the line of sight. In extreme cases, $\chi(\gamma = 0^\circ)$ is simply the density-weighted mean of the position angle $\psi$ of $\mathbf{B}_\perp$ along the line of sight, and if the plane-of-sky magnetic field is uniform (constant $\psi$) along the line of sight, $p \approx p_0\cos^2\gamma$. Any increase in $\gamma$ or in the dispersion level of $\psi$ along the line of sight would result in a lower polarization fraction. Here we discuss a mathematical analysis to quantify these two effects.

The position angle $\psi$ is defined as the angle (clockwise) between $\hat y$ and $\mathbf{B}_\perp$ in a convention that $0<\psi<\pi$ (see Figure~\ref{angDef}). We thus define the dispersion of the direction of $\mathbf{B}_\perp$ along the line-of-sight ($\hat z$) as the density-weighted standard deviation of $\psi$ along $\hat z$:
\begin{equation}
{\cal S}_\mathrm{los} \equiv \sqrt{\frac{\sum_{z} n \cdot(\psi - \overline{\psi})^2}{\sum_\mathrm{z} n}},
\end{equation}
where $\overline{\psi}$ is the density-weighted mean of $\psi$ along the line of sight:
\begin{equation}
\overline{\psi} \equiv \frac{\sum_z n \cdot \psi}{\sum_z n}.
\end{equation}
The parameter ${\cal S}_\mathrm{los}$ is a measurement of the level of tangledness of the plane-of-sky magnetic field along the line of sight.

We can also calculate the density-weighted average inclination angle with respect to the sky as
\begin{equation}
\langle \cos^2\gamma\rangle_\mathrm{los} \equiv \frac{\sum_z n\cdot \cos^2\gamma}{\sum_z n},
\label{cos2gdef}
\end{equation}
and
\begin{equation}
\langle \gamma\rangle_\mathrm{los} \equiv \arccos \left(\sqrt{\langle \cos^2\gamma\rangle_\mathrm{los} }\right).
\end{equation}
This is roughly the average inclination angle of the magnetic field with respect to the plane of sky along the line of sight. Note that $0<\langle \gamma\rangle_\mathrm{los} < \pi/2$.
These two parameters, ${\cal S}_\mathrm{los}$ and $\langle\cos^2\gamma\rangle_\mathrm{los}$, can be considered as two independent factors controlling the simulated polarization fraction at each pixel of the 2D projected map.

\begin{figure*}
\begin{center}
\includegraphics[width=0.9\textwidth]{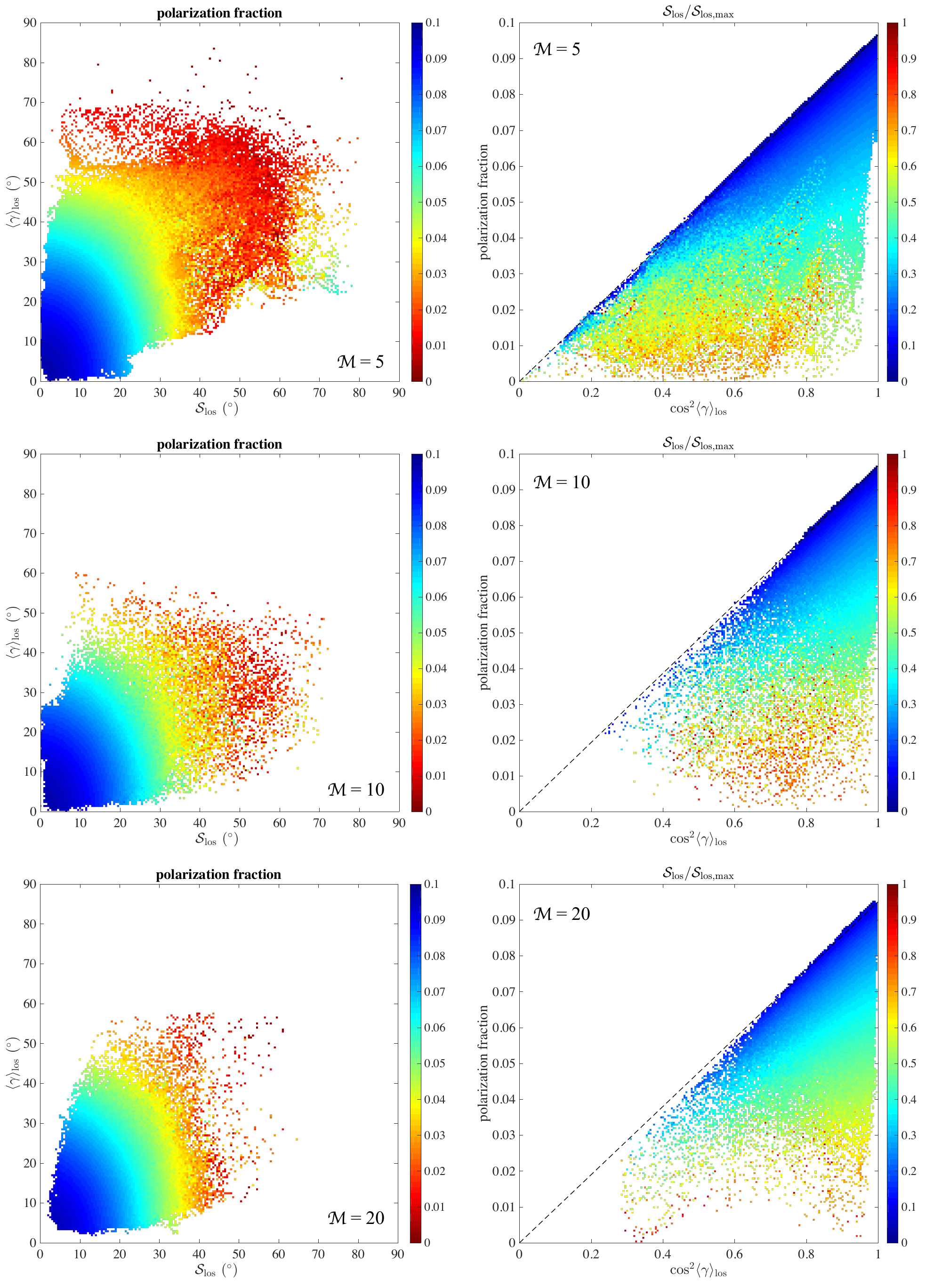}
\caption{The relationship between polarization fraction and the two controlling factors: inclination ($\langle\gamma\rangle_\mathrm{los}$) and tangledness (${\cal S}_\mathrm{los}$), for models with inflow Mach number ${\cal M} = 5$ ({\it top}), $10$ ({\it middle}), and $20$ ({\it bottom}). {\it Left:} color-coded polarization fraction $p$ of mean inclination angle $\langle\gamma\rangle_\mathrm{los}$ against plane-of-sky dispersion ${\cal S}_\mathrm{los}$. {\it Right:} color-coded normalized ${\cal S}_\mathrm{los}$ of $p$ against inclination factor $\langle\cos^2\gamma\rangle_\mathrm{los}$. }
\label{polarization}
\end{center}
\end{figure*}

The scatter plots of $\langle\gamma\rangle_\mathrm{los}$ and ${\cal S}_\mathrm{los}$ from the three simulation models considered in this study are shown in Figure~\ref{polarization} ({\it left panel}). These plots are color-coded by the polarization fraction; the dependence of $p$ on both ${\cal S}_\mathrm{los}$ and $\langle\gamma\rangle_\mathrm{los}$ is clear. Also, the distribution of the polarization fraction (shown in color) is very similar across the three models. 
This demonstrates that ${\cal S}_\mathrm{los}$ and $\langle\cos^2\gamma\rangle_\mathrm{los}$ are good indicators of the polarization fraction. 
The most striking feature of these plots is that the equation of an ellipse centered at the origin of ${\cal S}_\mathrm{los}$ and $\langle\gamma\rangle_\mathrm{los}$ seems to be a good fit for every single value of $p$; i.e.~the distribution of $\langle\gamma\rangle_\mathrm{los}$ vs. ${\cal S}_\mathrm{los}$ at a given value of $p$ can be roughly described by ${\langle\gamma\rangle_\mathrm{los}}^2/{a_1(p)}^2 + {{\cal S}_\mathrm{los}}^2/{a_2(p)}^2 = 1$, where $a_1(p)$ and $a_2(p)$ are constant coefficients that only depend on $p$.
However, we remind the readers that, with the real value of $p_0$ unknown, the polarization fraction derived in this study is only an approximation.
We therefore focus not on making quantitative predictions but instead on understanding physically the behavior of the polarization fraction. 

Figure~\ref{polarization} ({\it left panel}) suggests that the tangledness and inclination are almost equally important on determining the polarization fraction. To investigate their effects separately, we plotted the polarization fraction as a function of $\langle\cos^2\gamma\rangle_\mathrm{los}$ in Figure~\ref{polarization} ({\it right panel}), color-coded by normalized ${\cal S}_\mathrm{los}$ ($\equiv {\cal S}_\mathrm{los}/{\cal S}_\mathrm{los,max}$ in each model). 
One prominent feature among all three models is that $p \leq p_\mathrm{max}\cdot\langle\cos^2\gamma\rangle_\mathrm{los}$ (where $p_\mathrm{max}$ is the maximum value of $p$ measured in each model), a relation that can be proved mathematically (see Appendix~\ref{gammader}). 

More importantly, in Figure~\ref{polarization} ({\it right panel}) all points along the upper-boundary line $p = p_\mathrm{max}\cdot\langle\cos^2\gamma\rangle_\mathrm{los}$ ({\it dashed lines}) have normalized ${\cal S}_\mathrm{los} \ll 1$, and ${\cal S}_\mathrm{los}$ increases away from the line.
The color distribution suggests that at each point on the plane of sky, the averaged inclination angle of the magnetic field along the line of sight first gives the upper limit of the value $p$, then the tangledness of $\mathbf{B}_\perp$ along the line of sight determines how far away from this maximum value the polarization fraction is.

Note that a similar analysis has been done by \cite{2015A&A...576A.105P}, who also found the proportionality of the envelope of $p$ with $\langle\cos^2\gamma\rangle_\mathrm{los}$ (see their Figure~21). Though the range of gas density in their simulations is not wide enough to cover the self-gravitating regime where $B_\parallel/B$ dramatically increases (smallest $\langle\cos^2\gamma\rangle_\mathrm{los}$ values), by observing the simulation box at different viewing angles (denoted as $\alpha$ in their work) they were able to find the similar dependence of $p$ on $\langle\cos^2\gamma\rangle_\mathrm{los}$ as illustrated in Figure~\ref{polarization} ({\it right panel}). Also, their Equation~(9) can be adopted to give the upper-boundary line $p = p_\mathrm{max}\cdot\langle\cos^2\gamma\rangle_\mathrm{los}$ considering the viewing angle is approximately the inclination angle of the magnetic field in the case that $\mathbf{B} \approx \mathbf{B}_\perp$ (also see their Figure~16).

\subsection{Cloud Evolution Implied by Polarization Fraction}
\label{sec:evo}

\begin{figure*}
\begin{center}
\includegraphics[width=\textwidth]{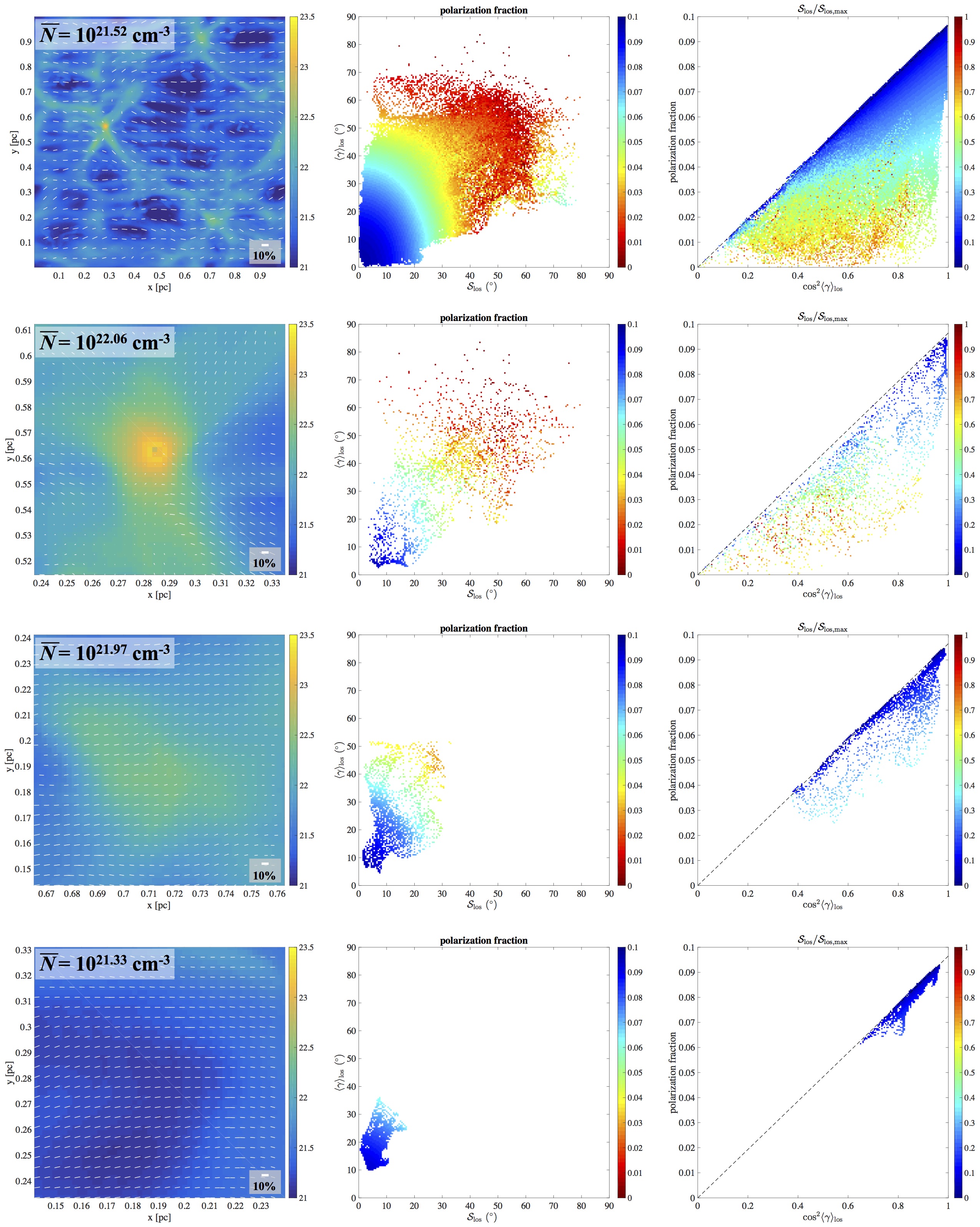}
\caption{Full map ({\it top row}) and sub-regions at different evolutionary stages ({\it rows 2, 3, and 4}) from model M5. {\it Left:} projected column density ({\it colormap}) with integrated polarization fraction and direction ({\it white segments}). {\it Middle and right:} Same as Figure~\ref{polarization}, but correspond to different regions of the map, as shown in the left column.}
\label{regionsM5}
\end{center}
\end{figure*}

\begin{figure*}
\begin{center}
\includegraphics[width=\textwidth]{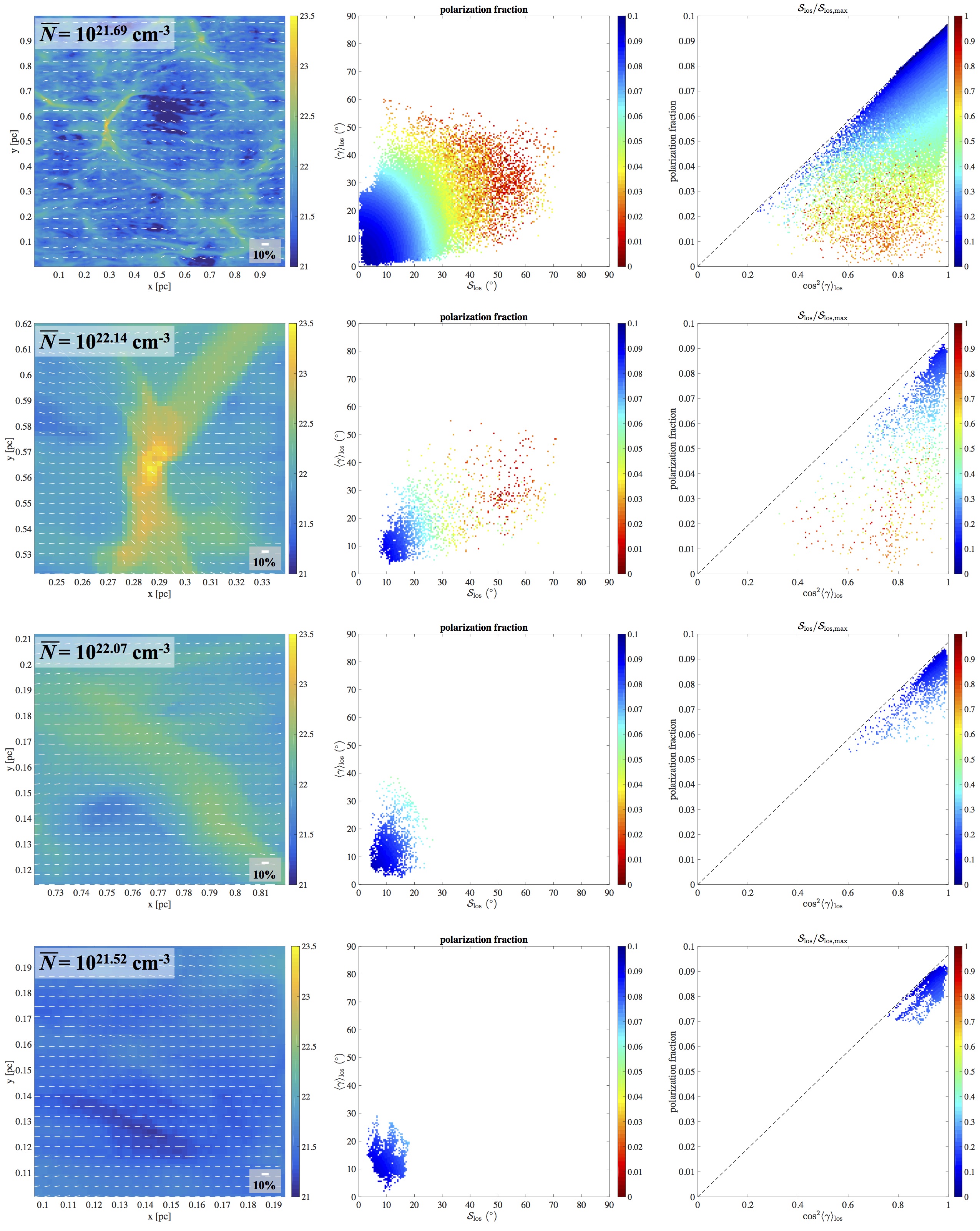}
\caption{Same as Figure~\ref{regionsM5}, but from model M10.}
\label{regionsM10}
\end{center}
\end{figure*}

\begin{figure*}
\begin{center}
\includegraphics[width=\textwidth]{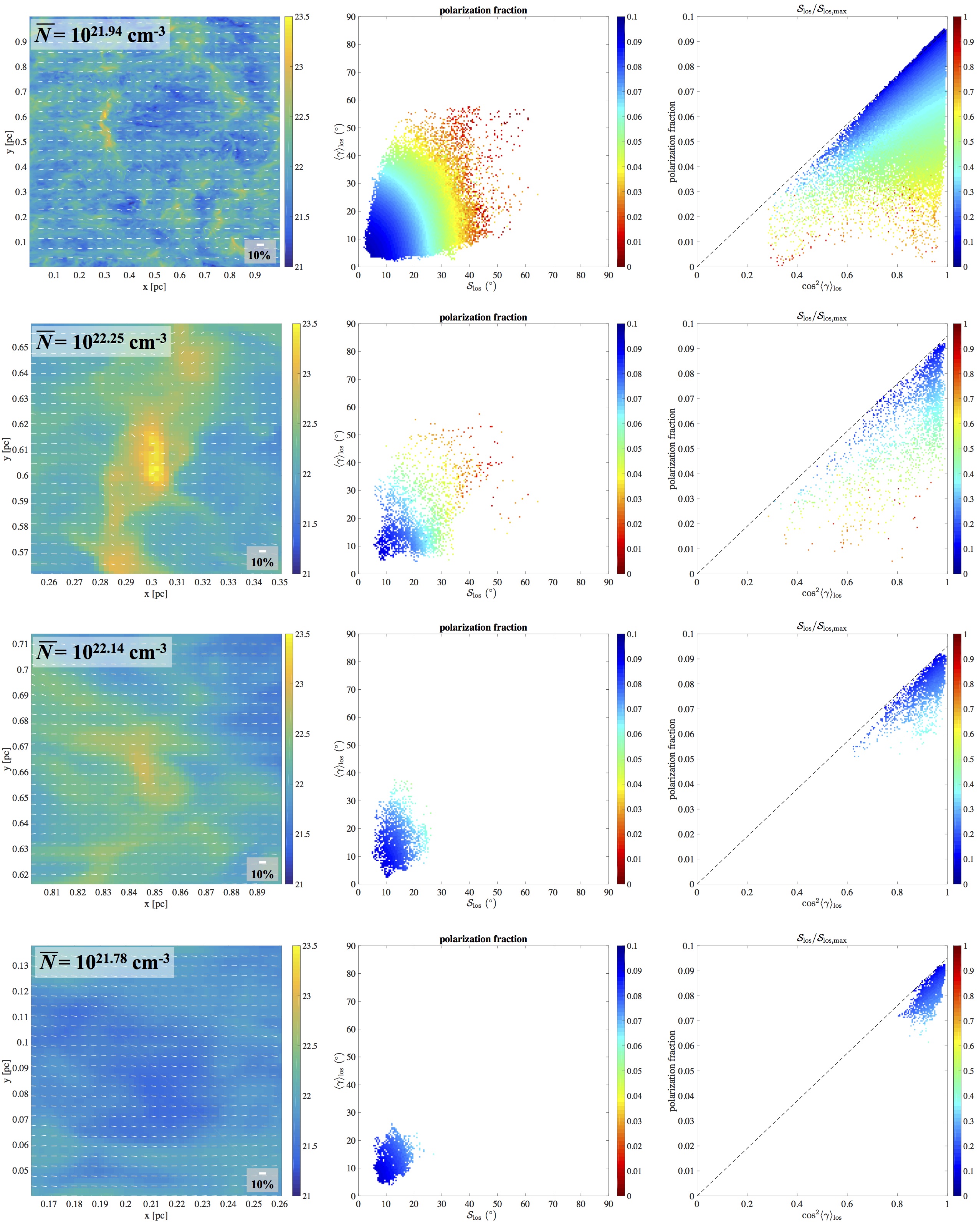}
\caption{Same as Figure~\ref{regionsM5}, but from model M20.}
\label{regionsM20}
\end{center}
\end{figure*}

We have already seen that polarization fraction generally decreases with increasing column density (see Section~\ref{sec:pvsN} and Figure~\ref{pvsN_Ma}).
To understand how polarization fraction is related to structure evolution and gas dynamics in the cloud, we selected regions from the projected map with different mean column densities $\overline{N}$, and plotted the relations between $p$, ${\cal S}_\mathrm{los}$, and $\langle\gamma\rangle_\mathrm{los}$ as in Figure~\ref{polarization} within these regions.
Figures~\ref{regionsM5} to \ref{regionsM20} are results from model M5, M10, and M20, respectively. For each model, in addition to the whole map ({\it top row}), we picked a core-forming region (high $\overline{N}$; {\it second row}), a filamentary region (intermediate $\overline{N}$; {\it third row}), and a region with no prominent structures (background/low $\overline{N}$; {\it bottom row}). In each figure, we showed the column density map with polarization segments of the region we selected ({\it left}), the $\langle\gamma\rangle_\mathrm{los} - {\cal S}_\mathrm{los}$ scatter plot color-coded by $p$ ({\it middle}), and the $p-\langle\cos^2\gamma\rangle_\mathrm{los}$ relation color-coded by normalized ${\cal S}_\mathrm{los}$ ({\it right}).

For all three models, the core-forming regions have the largest scatter ranges of ${\cal S}_\mathrm{los}$ and $\langle\gamma\rangle_\mathrm{los}$. As the mean column density decreases, both ${\cal S}_\mathrm{los}$ and $\langle\gamma\rangle_\mathrm{los}$ decrease. However, the scatter range of ${\cal S}_\mathrm{los}$ shrinks much faster than $\langle\gamma\rangle_\mathrm{los}$ (see the middle columns of Figures~\ref{regionsM5} to \ref{regionsM20}); also, in the less dense regions, points are preferentially distributed along the $p=p_\mathrm{max}\cdot\langle\cos^2\gamma\rangle$ lines (Figures~\ref{regionsM5} to \ref{regionsM20}, {\it right}). This implies that in the least dense regions which resemble the background environment of the cloud, the variation of $p$ is mainly induced by the inclination of magnetic field with respect to the plane of sky. It is not until the formation of dense clumps (which may be self-gravitating) that the tangledness of plane-of-sky magnetic field starts to set in and further reduces the polarization fraction.

Combined with the anisotropic condensation model described in \hyperlink{CO14}{CO14} and \hyperlink{CO15}{CO15}, our results suggest a three-stage development process of the magnetic field during the evolution of star-forming regions in molecular clouds. First, the initial magnetic field in the 
diffuse cloud ($B_0\sim 10~\mu$G) is amplified by a large-scale shock compression. In the post-shock layer where the magnetic field is strong and well ordered,  materials preferably flow along the field lines to form filamentary structures, which leaves $\mathbf{B}_\perp$ unchanged but may increase $\mathbf{B}_\parallel$ as well as the inclination angle $\gamma$. After dense clumps and/or prestellar cores gathered enough mass and become self-gravitating, the contraction/collapse induced by self-gravity is more isotropic. The magnetic field is thus amplified again ($B_\mathrm{core}\sim 10^2~\mu$G), which leads to higher levels of tangledness in $\mathbf{B}_\perp$ and lower values of $p$.

Note that the evolution of magnetic field can also explain the $p-N$ relation described in Section~\ref{sec:pvsN}. The anisotropic condensation stage is responsible for the power-law relation $p\propto N^{-\alpha}$ at $N < N_\mathrm{tr}$, because gas flows along $\mathbf{B}_\perp$ in the post-shock layer (which generate structures with higher column density) increase the inclination angle $\gamma$ and therefore reduce the polarization fraction. This is also why models with higher inflow Mach number has shallower $p-N$ curve (smaller $\alpha$; see Table~\ref{sumtable}), because stronger shocks create post-shock layers with smaller $B_\parallel / B_\perp$, and therefore $\gamma$ remains small even after $B_\parallel$ being amplified by the anisotropic condensation along $\mathbf{B}_\perp$. In dense regions where self-gravity dominates ($N > N_\mathrm{tr}$ regime), the dominant factor for reducing polarization fraction switches from inclination to tangledness. Since the effect of tangledness on the polarization fraction is not monotonic (the integration of $\sin 2\psi$ and $\cos 2\psi$ can lead to either cancellation or addition), the $p-N$ relation is relatively random at high column densities.

\subsection{Link to Observations}

From Figure~\ref{polarization}, for each value of polarization fraction measured in observation (i.e.~for every pixel of the 2D map), we can estimate the maximum value of averaged inclination angle $\langle\gamma\rangle_\mathrm{los}$ of the magnetic field. 
Furthermore, our results indicate that if any two of the three quantities $p$, ${\cal S}_\mathrm{los}$, and $\langle\cos^2\gamma\rangle_\mathrm{los}$  are known, the third one can be derived.
Unfortunately, neither the inclination angle nor the magnetic field structure along the line of sight is observable. However, the dispersion of polarization angle on the plane of sky, ${\cal S}_\mathrm{pos}$, can be derived directly from the measured polarization emission \citep[e.g.][]{2016ApJ...824..134F}. 
Though there is no priori physical reason for the magnetic field structure to be similar along the line of sight and over the plane of sky, it is reasonable for the dispersions of $\mathbf{B}_\perp$ in these two directions, ${\cal S}_\mathrm{los}$ and ${\cal S}_\mathrm{pos}$, to be correlated, since they are both induced by self-gravity and thus are dominated by the densest structures.

\begin{figure}
\begin{center}
\includegraphics[width=0.9\columnwidth]{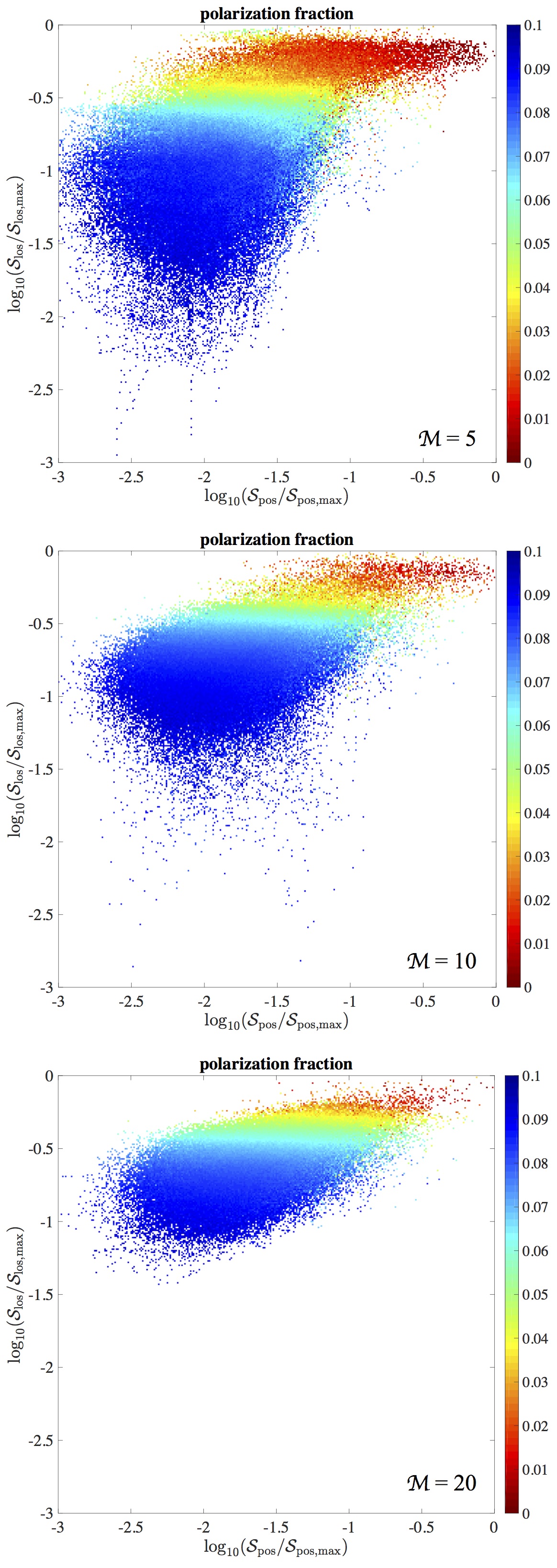}
\caption{The dispersion of plane-of-sky magnetic field along the line of sight, ${\cal S}_\mathrm{los}$, scatter-plotted against the dispersion on the plane of sky, ${\cal S}_\mathrm{pos}$. Both axes are normalized by the maximum value in each model M5 ({\it top}), M10 ({\it middle}), and M20 ({\it bottom}). The distributions are color-coded by the polarization fraction at each corresponding pixel.}
\label{SPOS}
\end{center}
\end{figure}

We calculated ${\cal S}_\mathrm{pos}$ following the definition
\begin{equation}
{{\cal S}_\mathrm{pos}}^2\big|_{(i,j)} = \frac{1}{8}\sum_{(i', j')} \left[\chi(i, j) - \chi(i', j')\right]^2
\end{equation}
where $(i', j')$ range from $(i-1,j-1)$ to $(i+1, j+1)$ so that ${\cal S}_\mathrm{pos}$ represents the dispersion of polarization angle $\chi$ at each pixel among its $8$ nearest neighbors. 
Figure~\ref{SPOS} illustrates a positive correlation between ${\cal S}_\mathrm{los}$ and ${\cal S}_\mathrm{pos}$, as expected. Therefore, for each value of ${\cal S}_\mathrm{pos}$ derived from the observed polarization map, the range of ${\cal S}_\mathrm{los}$ can be estimated, which (combined with the observed polarization fraction) can provide a hint of the 3D magnetic field morphology by the corresponding $\langle\gamma\rangle_\mathrm{los}$ value from Figure~\ref{polarization}. 
Another interesting feature revealed in Figure~\ref{SPOS} is that the polarization fraction ({\it colormap}) is nearly independent of ${\cal S}_\mathrm{pos}$ at a given ${\cal S}_\mathrm{los}$ value. Also, the distribution of a given value of $p$ in Figure~\ref{SPOS} can be roughly described as $\log {\cal S}_\mathrm{los} = b_1(p) \log {\cal S}_\mathrm{pos} + b_2(p)$, which provides a more straightforward way to estimate ${\cal S}_\mathrm{los}$ from observations.
More detailed discussion about quantitative derivation of polarization-implied magnetic field morphology is postponed to a follow-up paper (King et al., {\it in prep.}).



\section{Summary}
\label{sec:sum}

Using the Histogram of Relative Orientation (HRO) introduced by \cite{2013ApJ...774..128S}, we investigated the gas density and magnetic field structure in MHD simulations of prestellar core-forming regions formed through shock compression in GMCs, as described in \hyperlink{CO14}{CO14} and \hyperlink{CO15}{CO15}. We found that the relative orientation between density structure and magnetic field is directly impacted by the importance of self-gravity in the post-shock layer. Specifically, the magnetic field morphology changes in dense regions where the gas is accelerated gravitationally and becomes super-Alfv{\'e}nic. The HRO shape changes from concave to convex at the transition density $n_\mathrm{tr}$, which is also the density for the field strength to transition from a more or less constant to a power-law increase with density. In the projected 2D map, a similar change in HRO shape is found to correlate with the column density below which the polarization fraction varies monotonically with column density. We interpreted the decrease of polarization fraction in terms of enhanced tangledness and inclination of the magnetic field, which, together with HROs, may be applied to polarization observations to provide estimates on the total magnetic field strength and the 3D field structure.

Our main conclusions are as follows: 
\begin{enumerate}

\item
Our magnetic field$-$gas density plot shows characteristic features for the pre-shock, the post-shock, and the self-gravitating regions (Figure~\ref{BvsD_Ma}, {\it left panel}). The relatively unchanged magnetic field strength in the post-shock region ($n\sim 10^4-10^5$~cm$^{-3}$) is  direct evidence of the anisotropic gas flow along the magnetic field line (\hyperlink{CO14}{CO14}, \hyperlink{CO15}{CO15}). In the super-Alfvenic, self-gravitating regime ($n > n_\mathrm{tr} \sim 10^5$~cm$^{-3}$), the magnetic field strength is strongly correlated with the gas density with a power-law scaling (see below).

\item 
For densities above the threshold value $n_\mathrm{tr}$, the $B-n$ curve follows a power law $B \propto n^\kappa$ (blue dotted lines in Figure~\ref{BvsD_Ma}, {\it left panel}) with $\kappa\approx 0.3-0.6$ (Table~\ref{sumtable}). The values of $\kappa$ found in this work are consistent with previous observational and theoretical studies \citep[e.g.][]{1999ApJ...520..706C,2001ApJ...546..980O,2010ApJ...725..466C,2014MNRAS.437...77B,2015MNRAS.452.2500L}. Our fitting results imply that the field$-$density relation has strong dependence on the turbulent level of the cloud, with smaller $\kappa$ for weaker turbulence.  (Section~\ref{sec:bn}).

\item
We found that the threshold density in the $B-n$ relation, $n_\mathrm{tr}$, is also where the 3D HRO shape changes from concave to convex. In all three models considered in this work, HROs with segmentations $n < n_\mathrm{tr}$ are concave, while those with segmentations $n > n_\mathrm{tr}$ are either flat or convex (Figure~\ref{BvsD_Ma}, {\it right panel}; also see Figure~\ref{Bn_zeta}). This suggests that the shape of HRO, or the magnetic field$-$gas structure alignment, is strongly altered by the self-gravity of the gas in the high-density regime. 
Specifically, the HRO changes shape when the sub-Afv\'enic gas is gravitationally accelerated to a super-Afv\'enic speed, so that ${\cal M}_\mathrm{A} \approx 1$ at $n\approx n_\mathrm{tr}$ (Figures~\ref{energy} and \ref{MA}). The HRO can therefore be used, at least conceptually, to estimate the density at the Alfv\'enic transition, which, together with velocity information, yields the total field strength. 

\item
Similar behavior of HRO in the 3D space is found in the 2D projected map, between the integrated polarization pseudo-vector $\boldsymbol{p}$ (see Equation~(\ref{pvec}) for definition) and the gradient of column density $N$ (Figure~\ref{pvsN_Ma}). The transition of HRO shape from concave to convex is related to the threshold value for the $p-N$ curve to stay relatively monotonic, $N_\mathrm{tr}$ (Figures~\ref{pvsN_Ma} and \ref{pvsN_zeta}). 
Our analysis showed that when the gas content is self-gravitating ($n > n_\mathrm{tr}$), the magnetic field morphology is significantly altered (Figure~\ref{dcutmap}), and therefore the integrated polarization fraction along the line-of-sight is no longer dominated by the gas column density. As a result, the change of HRO shape in the 2D projected map is also a consequence of increasing significance of gas self-gravity, and can be directly linked to the threshold density $n_\mathrm{tr}$ (Figure~\ref{dcutmap}).

\item
For the non-self-gravitating regime ($N < N_\mathrm{tr}$) in the 2D projected map, we found and fitted a power-law relationship to the $p-N$ curve (blue dotted lines in Figure~\ref{pvsN_Ma}, {\it left panel}), $p \propto N^{-\alpha}$, with the index $\alpha \sim 0.0-0.4$ (Table~\ref{sumtable}). These values are roughly consistent with, but at the low end of, previous observational and theoretical studies \citep[e.g.][]{2001ApJ...562..400M,2005A&A...430..979G,2008ApJ...679..537F,2015A&A...576A.105P,2016ApJ...824..134F}. Based on our results, larger value of $\alpha$ may indicate a less turbulent environment, or a relatively weakly-magnetized medium (Table~\ref{sumtable}). Further simulations with more extended model parameters are necessary to firm up the conclusion. 

\item 
We demonstrated the polarization fraction is controlled by two factors: 1) the dispersion of the plane-of-sky magnetic field $\mathbf{B}_\perp$ along the line-of-sight, ${\cal S}_\mathrm{los}$, and 2) the averaged inclination angle of the magnetic field along the line-of-sight, $\langle\gamma\rangle_\mathrm{los}$ (Figure~\ref{polarization}, {\it left panel}). In addition, we showed that the mean inclination sets up an upper limit for the integrated polarization fraction (Appendix~\ref{gammader}), while the tangledness of $\mathbf{B}_\perp$ determines the distance of the actually measured polarization fraction to that maximum value (Figure~\ref{polarization}, {\it right panel}). Combined with the fact that there is a positive correlation between the dispersion of polarization angle on the plane of sky ${\cal S}_\mathrm{pos}$ and the line-of-sight dispersion ${\cal S}_\mathrm{los}$ (Figure~\ref{SPOS}), the 3D magnetic field morphology can be constrained from the observed polarization fraction and the derived dispersion (King~et al., {\it in prep.}).

\item
We established an evolutionary model of core-forming regions in GMCs for the magnetic field morphology (Section~\ref{sec:evo}). In the diffuse cloud, the large-scale turbulence creates shocked regions with a relatively well-ordered magnetic field parallel to the shock front (\hyperlink{CO14}{CO14}, \hyperlink{CO15}{CO15}). When observed perpendicular to the shock front, this cloud background has polarization fraction $p = p_\mathrm{max} \sim 10\%$, with $\langle\gamma\rangle_\mathrm{los} \approx 0^\circ$ and ${\cal S}_\mathrm{los} \approx 0^\circ$ (Figures~\ref{regionsM5}$-$\ref{regionsM20}, {\it bottom panel}). Within this sub-Alfv{\'e}nic (Figures~\ref{energy}$-$\ref{MA}), relatively-flat (Figures~3$-$5 in \hyperlink{CO14}{CO14}) post-shock region, gas flows preferentially along the magnetic field lines to form dense structures (\hyperlink{CO14}{CO14}, \hyperlink{CO15}{CO15}). These anisotropic flows will gradually compress the line-of-sight component of the magnetic field; therefore $\langle\gamma\rangle_\mathrm{los}$ increases and $p$ decreases (Figures~\ref{regionsM5}$-$\ref{regionsM20}, {\it third panel}). After a self-gravitating prestellar core is formed, the gravity-induced isotropic contraction in both gas and magnetic field will enhance the tangledness of magnetic field and further decrease the polarization fraction (Figures~\ref{regionsM5}$-$\ref{regionsM20}, {\it second panel}).

\end{enumerate}

To conclude, the success of using the transition density in $B-n$ relation to explain different shapes of HRO in both 3D simulation domain and 2D projected map is encouraging, which also suggests links between observable quantities ($N$, $E_\mathrm{K}$, $p$, ${\cal S}_\mathrm{pos}$) and gas properties in real space ($n_\mathrm{tr}$, $E_\mathrm{B}$, ${\cal S}_\mathrm{los}$, $\langle\gamma\rangle_\mathrm{los}$). This provides strong motivation for testing these ideas in observations of various density tracers and of polarized thermal continuum. 

\acknowledgements
We thank the anonymous referee for a very informative report.
C.-Y.~C. is grateful for the support from Virginia Institute of Theoretical Astronomy (VITA) at the University of Virginia through the VITA Postdoctoral Prize Fellowship. Z.-Y.~L. is supported in part by NSF AST-1313083 and NASA NNX14AB38G. P. K.~K. is supported by the National Radio Astronomy Observatory through a Student Observing Support grant and by the Jefferson Scholars Foundation at the University of Virginia through a graduate fellowship. 

\appendix

\section{Maximum Polarization Fraction and Magnetic Field Inclination Angle}
\label{gammader}

The relationship between the relative Stokes parameters and the observables, $N$, $p$, and $\chi$ can be expressed in a compact way using complex algebra. If we define the \textit{normalized polarization fraction} as 

\begin{equation}
\bar{p} = \frac{p}{p_0}
\end{equation}

\noindent and the \textit{inclination-corrected column density} as

\begin{equation}
\bar{N} = N - p_0 N_2,
\label{corNdef}
\end{equation}

\noindent then 

\begin{equation}
\bar{p}\bar{N} = \sqrt{q^2 + u^2} = \text{mod}(q + i u),
\end{equation}
\begin{equation}
2\chi = \arctan2(u,q) = \text{arg}(q + i u),
\end{equation}

\noindent and so our compact relation is (see Equation~\ref{poldef2})

\begin{equation}
\bar{p}\bar{N} e^{2i\chi} = q + i u = \int n \frac{B_y^2 + 2iB_xB_y - B_x^2}{B^2}~ ds = \int n \frac{(B_y + iB_x)^2}{B^2} ~ds.
\end{equation}

Noting that $\bar{p}$ and $\bar{N}$ are both positive-semidefinite, 

\begin{equation}
\bar{p}\bar{N} = \left|\int n \frac{(B_y + iB_x)^2}{B^2} ~ds \right|.
\end{equation}

\noindent Utilizing the triangle inequality for complex numbers and the fact that $\rho$ and $B^2$ are positive-semidefinite, we have

\begin{equation}
\bar{p}\bar{N} \leq \int n \frac{B_x^2 + B_y^2}{B^2} ds,
\end{equation}

\noindent which is nothing more than the density-weighted integral (along the line of sight) of $\cos^2 \gamma$: 

\begin{equation}
\bar{p} \bar{N} \leq \int n \cos^2 \gamma ~ds. 
\end{equation}

\noindent Recalling our definitions for $\bar{N}$ (Equation~(\ref{corNdef})) and $\langle \cos^2 \gamma \rangle$ (Equation~(\ref{cos2gdef})),

\begin{equation}
\bar{p} \left(1 - p_0 \frac{N_2}{N}\right) \leq \langle \cos^2 \gamma \rangle,
\end{equation}
\begin{equation}
\bar{p} \left( 1 + \frac{p_0}{3} - \frac{p_0}{2} \langle\cos^2 \gamma \rangle\right) \leq \langle \cos^2 \gamma \rangle,
\end{equation}

\noindent which gives us the exact result

\begin{equation}
\frac{p}{p_0} \leq \frac{\langle \cos^2 \gamma \rangle}{1 + \frac{p_0}{3} - \frac{p_0}{2}\langle \cos^2 \gamma \rangle},
\end{equation}

\noindent and in turn, for $p_0 \ll 1$, gives the compact approximate relation

\begin{equation}
\frac{p}{p_0} \lesssim \langle \cos^2 \gamma \rangle.
\end{equation}

We recover that the (normalized) polarization fraction is reduced with $\langle \cos^2 \gamma \rangle$ - that is, lines of sight with a higher degree of magnetic inclination will have lower polarization fractions. This also implies that, for a given line of sight, the polarization fraction is maximized only when two conditions are met: that the magnetic field orientation on the plane of the sky has low variation (the field is ordered) and that the magnetic inclination is minimized (the magnetic field is principally in the plane of the sky). 
As noted above, this derivation does not use the revised definition of $N_2$ claimed by \cite{2015A&A...576A.105P}. The derivation can be simply modified to accommodate the new definition.

\end{document}